\documentclass[preprint,12pt]{elsarticle}

\usepackage{amssymb}
\usepackage{amsmath}
\usepackage{siunitx}
\usepackage{graphicx}
\usepackage{subcaption}
\usepackage{hyperref}
\usepackage{booktabs}
\usepackage{multirow}
\usepackage{makecell}
\usepackage{tikz}
\usepackage{hyphenat}
\usepackage{here}

\journal{Nucl. Instrum. Methods A}

\begin{document}

\begin{frontmatter}



\title{Construction and performance evaluation of a meter-scale monolithic scintillator tracker with embedded scatterers}


\author[a]{N.~Otani\corref{cor1}}
\ead{otani.naoki.53x@st.kyoto-u.ac.jp}
\author[a]{S.~Han}
\author[b]{E.~Abad~D\'{i}az}
\author[c]{S.~Aoki}
\author[b]{D.~Bar\v{c}ot}
\author[d]{C.~Bronner}
\author[e]{T.~Fukuda}
\author[b]{M.~Ghosh}
\author[b]{L.~Hali\'{c}}
\author[a]{K.~Hasegawa}
\author[e]{T.~Hayakawa}
\author[e]{Y.~Hayasaka}
\author[a]{K.~Hayashi}
\author[d]{D.~Hirata}
\author[a]{K.~Indo}
\author[b]{M.~Jakkapu}
\author[e]{A.~Kasumi}
\author[f]{T.~Katori}
\author[e]{T.~Kawahara}
\author[a]{M.~Kawaue}
\author[a]{T.~Kikawa}
\author[b]{B.~Kli\v{c}ek}
\author[e]{H.~Kobayashi}
\author[b]{B.~Kova\v{c}}
\author[a]{J.~McKean}
\author[e]{T.~Matsuo}
\author[g]{S.~Mikado}
\author[d]{A.~Minamino\fnref{fn1}}
\author[e]{Y.~Morimoto}
\author[e]{Y.~Nakamura}
\author[a]{T.~Nakaya}
\author[e]{H.~Oaira}
\author[a]{N.~Onda}
\author[h]{H.~Oshima}
\author[e]{O.~Sato}
\author{(NINJA Collaboration)}

\affiliation[a]{organization={Department of Physics, Kyoto University},
            addressline={Kitashirakawa Oiwake-cho, Sakyo-ku},
            city={Kyoto},
            postcode={606-8502},
            country={Japan}}

\affiliation[b]{organization={Center of Excellence for Advanced Materials and Sensing Devices, Ru{\dj}er Bo\v{s}kovi\'{c} Institute},
            addressline={Bijeni\v{c}ka cesta 54},
            city={Zagreb},
            postcode={10000},
            country={Croatia}}

\affiliation[c]{organization={Kobe University},
            city={Kobe},
            postcode={657-8501},
            country={Japan}}

\affiliation[d]{organization={Department of Physics, Yokohama National University},
            addressline={79-5 Tokiwadai, Hodogaya-ku},
            city={Yokohama, Kanagawa},
            postcode={240-8501},
            country={Japan}}

\affiliation[e]{organization={Department of Physics, Nagoya University},
            addressline={Furo-cho, Chikusa-ku},
            city={Nagoya, Aichi},
            postcode={464-8602},
            country={Japan}}

\affiliation[f]{organization={Department of Physics, King's College London},
            addressline={Strand},
            city={London},
            postcode={WC2R 2LS},
            country={United Kingdom}}

\affiliation[g]{organization={Department of Liberal Arts and Basic Sciences, College of Industrial Technology, Nihon University},
            addressline={1-2-1 Izumi-cho},
            city={Narashino, Chiba},
            postcode={275-8575},
            country={Japan}}

\affiliation[h]{organization={School of Science and Engineering, Tokyo Denki University},
            addressline={Ishizaka, Hatoyama-machi, Hiki-gun},
            city={Saitama},
            postcode={350-0394},
            country={Japan}}

\cortext[cor1]{Corresponding author.}

\fntext[fn1]{Also at Institute of Advanced Sciences, Yokohama National University, 79-5 Tokiwadai, Hodogaya-ku, Yokohama, Kanagawa 240-8501, Japan}

\begin{abstract}

A meter-scale Fiber-Readout mOnolithic and Scatterer-\allowbreak embedded scintillator Tracker (FROST) was constructed and operated in the NINJA experiment to enable muon-track matching between the nuclear emulsion detectors and the downstream muon detector.
The detector consists of a $1.32~\mathrm{m}\times1.40~\mathrm{m}$ plastic scintillator plate assembled from optically bonded scatterer-embedded plates, with wavelength-shifting fibers arranged at a pitch of $10~\mathrm{mm}$.
During J-PARC neutrino-beam operation, FROST achieved a data acquisition efficiency of $99.59\%$ and a detection efficiency of $99.946\pm0.006\%$ for selected neutrino-induced muons well within the active area.
A position reconstruction method based on the light-yield distribution was developed for both single- and multi-particle events.
In a neutrino-interaction simulation, muon position resolutions of $1.77~\mathrm{mm}$ and $1.61~\mathrm{mm}$ were achieved in the two orthogonal readout directions, assuming correct muon identification.
Using a beam-data sample dominated by single-muon events, upper limits on the position resolution of $2.20~\mathrm{mm}$ and $2.03~\mathrm{mm}$ were obtained for track angles below $20^\circ$.
These results demonstrate that the FROST concept can be scaled up to a meter-scale active area while maintaining high detection efficiency and a position resolution well below the readout pitch under actual experimental conditions.

\end{abstract}


\end{frontmatter}


\section{Introduction}
\label{sec:introduction}

Plastic scintillators are widely used in tracking detectors because they provide fast response times and can be fabricated in various sizes to cover large active areas.
In conventional plastic scintillator trackers, the detector is typically divided into discrete scintillator elements, and the hit position is determined from the elements that are hit.
Using smaller segments improves the position resolution but increases the number of scintillator elements and readout channels.
This increase becomes particularly significant for large-area trackers, where it leads to higher cost and greater complexity of the readout system.
It is therefore important to achieve a position resolution finer than the readout pitch without requiring a corresponding increase in the number of readout channels when scaling scintillator trackers to large active areas.

To address this challenge, the Fiber-Readout mOnolithic and Scatterer-embedded scintillator Tracker (FROST) concept was proposed in a previous study~\cite{FROSTPrototype}.
FROST consists of a monolithic plastic scintillator plate with embedded scatterers and wavelength-shifting (WLS) fibers arranged in two orthogonal directions.
The scintillation light is collected by the WLS fibers and detected by silicon photomultipliers (SiPMs).
The collected light yield tends to be larger in channels whose fibers are closer to the charged-particle crossing position.
The embedded scatterers localize the scintillation light around the crossing position and further strengthen this tendency.
The crossing position is reconstructed from the channel-to-channel light-yield distribution with a position resolution well below the readout pitch.
The monolithic design also avoids inactive gaps associated with conventional scintillator segmentation, allowing near-100\% detection efficiency in principle.

The previous study demonstrated the basic performance of FROST using prototype detectors with $100~\mathrm{mm}\times100~\mathrm{mm}\times10~\mathrm{mm}$ scintillator plates and a readout pitch of $10~\mathrm{mm}$~\cite{FROSTPrototype}.
Scaling FROST to a meter-scale detector requires a method for constructing a large scintillator plate, appropriate treatment of the detector response over the larger active area, and stable readout during long-term operation.
In addition, the previous study focused on position reconstruction for single-particle events.
In realistic experimental environments, multiple charged particles may traverse the detector simultaneously, requiring a position reconstruction method that can handle multi-particle events.

In this work, multiple scatterer-embedded scintillator plates were optically bonded to realize a meter-scale active scintillator plate.
The resulting FROST detector, with an active area of $1.32~\mathrm{m}\times1.40~\mathrm{m}$, a readout pitch of $10~\mathrm{mm}$, and 272 readout channels, was constructed for the NINJA experiment and operated during the neutrino-beam exposure.
Using neutrino-induced muons recorded during this operation, its performance under actual experimental conditions was evaluated in terms of the detection efficiency and position resolution.

The remainder of this paper is organized as follows.
Section~\ref{sec:setup_requirements} introduces the NINJA experiment and the application of FROST, and Section~\ref{sec:designandcomponents} describes the design and components of the NINJA FROST detector.
Sections~\ref{sec:simulation} and~\ref{sec:reconstruction} present the optical simulation and position reconstruction, respectively.
Section~\ref{sec:operation} describes the detector operation during the neutrino-beam exposure, and Section~\ref{sec:performance} presents the performance evaluation using beam data.
Finally, Section~\ref{sec:conclusion} concludes the paper.

\section{Application of FROST to the NINJA experiment}
\label{sec:setup_requirements}

\subsection{The NINJA experiment}
\label{sec:ninja_motivation}

The NINJA (Neutrino Interaction research with Nuclear emulsion and J-PARC Accelerator) experiment aims to measure the kinematics of final-state particles in neutrino--nucleus interactions using nuclear emulsion detectors exposed to the muon-neutrino beam at J-PARC~\cite{NINJAPhysicsRun}.
The submicrometer position resolution of nuclear emulsion detectors enables the reconstruction of short charged-particle tracks, including protons with momenta down to approximately $200~\mathrm{MeV}/c$~\cite{NINJAPhysicsRun}.
Measurements of these final-state kinematics help constrain uncertainties in neutrino--nucleus interaction models, which constitute a major source of systematic uncertainty in neutrino oscillation measurements~\cite{T2KOscillation2023, NOvAOscillation2022}.
The NINJA experiment has conducted three physics runs, and this study uses the detector configuration from the third physics run, conducted from November 2025 to March 2026. For details of the detector operation during this run, see Section~\ref{sec:operation}.

\subsection{NINJA detector configuration and role of the scintillator tracker}
\label{sec:e71c_setup}

\begin{figure}[htbp]
  \centering
  \includegraphics[width=1.0\linewidth]{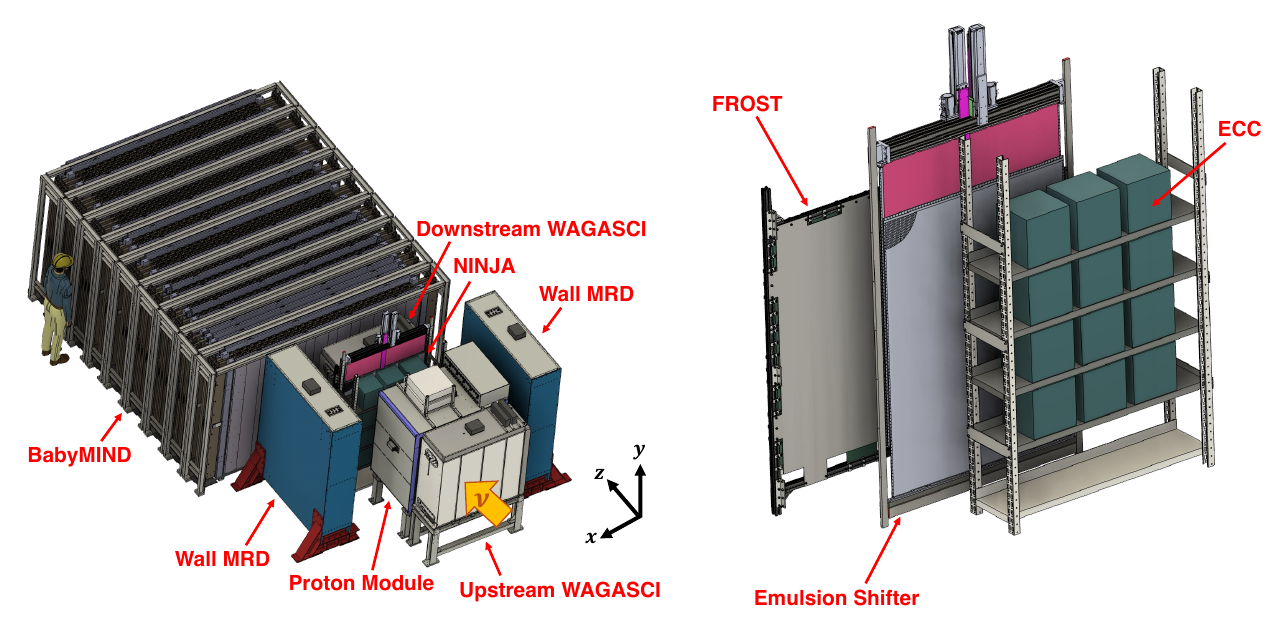}
  \caption{Placement of the NINJA detectors within the WAGASCI--BabyMIND detector setup (left) and an enlarged schematic view of the NINJA detectors used in this study (right).}
  \label{fig:ninja_wagasci_setup}
\end{figure}

The NINJA detectors were installed on the B2 floor of the Neutrino Monitor building located $280~\mathrm{m}$ downstream of the graphite target of the J-PARC neutrino beamline.
They were placed within the WAGASCI--BabyMIND detector setup~\cite{WAGASCIBabyMIND} of the T2K experiment, between the Proton Module and Downstream WAGASCI.
Figure~\ref{fig:ninja_wagasci_setup} shows their placement and the detailed detector configuration used in this study (the third NINJA physics run in 2025--2026).
The right-handed coordinate system used in this paper is defined with its origin at the center of the scintillator plate of the NINJA FROST detector.
The $z$ axis is defined as the direction normal to the scintillator plate, with the neutrino beam incident at an angle of approximately $5^\circ$ to the $z$ axis.
The $x$ and $y$ axes are the horizontal and vertical directions in the detector plane, respectively.

The Emulsion Cloud Chambers (ECCs), which served as both the neutrino interaction targets and tracking detectors, were located at the upstream end of the detector setup.
For the third NINJA physics run in 2025--2026 (this study's configuration), twelve ECCs were arranged in a $3 \times 4$ array in a plane perpendicular to the $z$ axis, and the total water target mass was increased from $74~\mathrm{kg}$ in the first two physics runs to $100~\mathrm{kg}$ to increase the neutrino-interaction statistics.
Each ECC consisted of alternating water-target layers and emulsion-based tracking layers~\cite{NINJAPhysicsRun}.

A muon range detector, BabyMIND (Magnetized Iron Neutrino Detector)~\cite{WAGASCIBabyMIND, BabyMIND}, which is part of the T2K near-detectors, is located at the downstream end of the detector setup.
BabyMIND consists of alternating layers of magnetized iron plates and plastic scintillator bars and measures the kinematics of muons emerging from the ECCs and identifies their charge sign from the direction of curvature in the magnetic field.

Nuclear emulsion provides submicrometer position resolution but has no intrinsic timing information and records all charged-particle tracks accumulated during the exposure.
In contrast, BabyMIND provides a timing resolution of $\mathcal{O}(\mathrm{ns})$, but its position resolution of $\mathcal{O}(\mathrm{cm})$ is insufficient to directly identify the corresponding muon track among the large number of tracks recorded in the ECCs.
To connect muon tracks between the ECCs and BabyMIND, an emulsion shifter and a scintillator tracker were used as intermediate detectors.
The emulsion shifter provides submicrometer position resolution and minute-to-hour-scale timing resolution by periodically shifting the nuclear emulsion films~\cite{NINJAPhysicsRun}.
The scintillator tracker provides position and timing resolutions of $\mathcal{O}(\mathrm{mm})$ and $\mathcal{O}(\mathrm{ns})$, respectively.

In the first two NINJA physics runs, a segmented scintillator tracker with an active area of $1.0~\mathrm{m}\times1.0~\mathrm{m}$ was used~\cite{E71aTracker}.
For the third physics run, a scintillator tracker with an active area of approximately $1.3~\mathrm{m}\times1.4~\mathrm{m}$ was required to cover the enlarged ECC array.
Increasing the active area with the same segmented design would require a corresponding increase in the number of scintillator bars and readout channels, resulting in higher cost and greater complexity of the readout system.
To enlarge the active area while limiting the increase in the number of readout channels, FROST was adopted as the scintillator tracker for the third physics run.

\subsection{Requirements for the NINJA FROST detector}
\label{sec:frost_requirements}

The NINJA FROST detector was required to provide sufficient position resolution for track matching, cover the enlarged ECC array, and maintain high detection efficiency for minimum-ionizing particles.
The required position resolution was set to $5.6~\mathrm{mm}$ to keep the accidental-matching probability between the scintillator tracker and the emulsion shifter below $5\%$, following the procedure used for the previous scintillator tracker~\cite{E71aTracker}.
An active area of approximately $1.3~\mathrm{m}\times1.4~\mathrm{m}$ was required to cover the enlarged ECC array.
Since inefficiency in muon-track matching between the detectors directly reduces the neutrino event yield, a muon detection efficiency close to 100\% was desired.

\section{Design and components of the NINJA FROST detector}
\label{sec:designandcomponents}

\subsection{Design}
\label{sec:design}

\begin{figure}[htbp]
  \centering
  \includegraphics[width=0.8\linewidth]{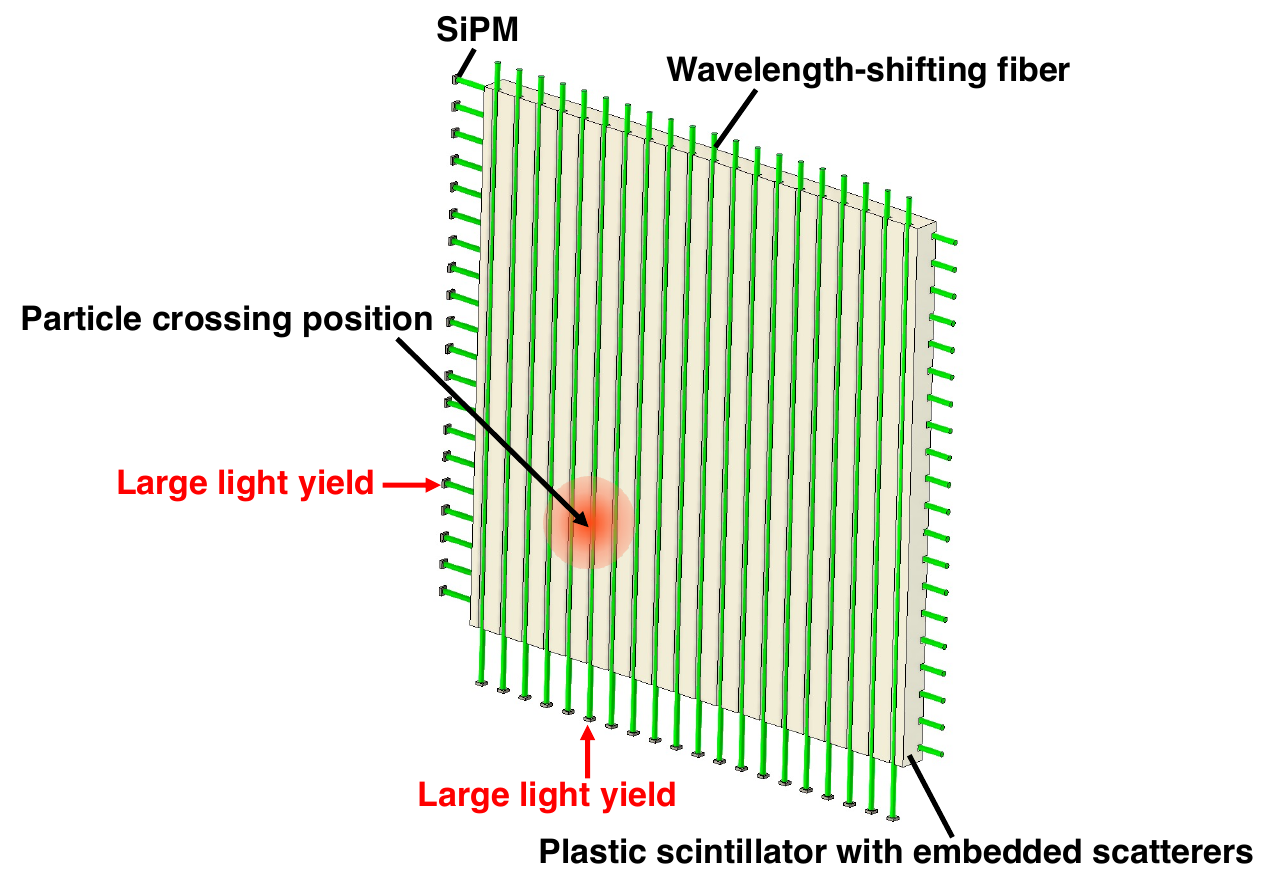}
  \caption{Conceptual design and position reconstruction principle of FROST. Only 20 fibers are shown in each direction for visual simplicity.}
\label{fig:frost_design}
\end{figure}

Figure~\ref{fig:frost_design} illustrates the conceptual design and position reconstruction principle of FROST.
FROST consists of a monolithic plastic scintillator plate with embedded scatterers and WLS fibers placed in evenly spaced grooves on the scintillator surfaces.
The fibers are placed on the two opposite surfaces in orthogonal directions to provide two-dimensional position information.
Each fiber is optically coupled to the scintillator and read out by a SiPM coupled to one end of the fiber.
The scintillator surfaces are coated with reflective paint to improve the light collection efficiency.

When a charged particle traverses the scintillator, it produces scintillation light, a fraction of which is collected by the WLS fibers.
The collected light yield tends to be larger in channels whose fibers are closer to the particle crossing position.
The embedded scatterers localize the scintillation light around the particle crossing position and further strengthen this tendency.
The particle position is reconstructed from the light-yield distribution across the readout channels, allowing a position resolution well below the readout pitch.

The NINJA FROST detector has an active area of $1.32~\mathrm{m}\times1.40~\mathrm{m}$.
The WLS fibers are arranged at a pitch of $10~\mathrm{mm}$, with 132 and 140 readout channels in the two directions, respectively, giving 272 channels in total.
The detailed detector components are described in the following subsections.

\subsection{Components}
\label{sec:components}

\subsubsection{Plastic scintillator}
\label{sec:plasticscintillator}

\begin{figure}[htbp]
  \centering
  \includegraphics[width=0.8\linewidth]{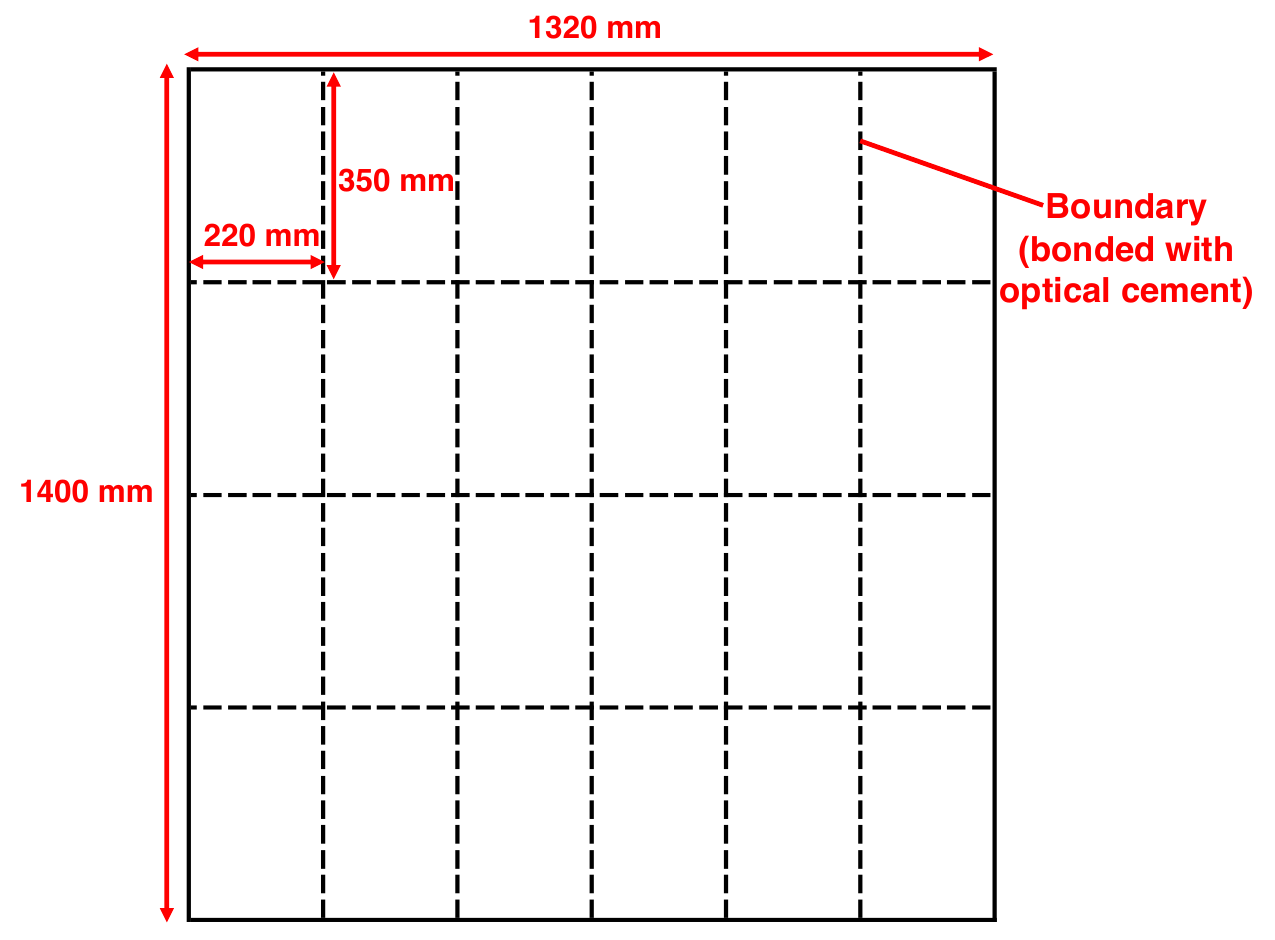}
  \caption{Schematic arrangement of the plastic scintillator plates forming the scintillator plate of the NINJA FROST detector.}
  \label{fig:scintillator_tiles}
\end{figure}

The scatterer-embedded plastic scintillator used in the NINJA FROST detector was custom manufactured by Kuraray to the same specifications as the scintillator used in Prototype M3 described in the previous study~\cite{FROSTPrototype}, which achieved the best position resolution among the prototypes evaluated.
To realize the required meter-scale active area, the scintillator plate was assembled from 24 smaller plates, each with dimensions of $220~\mathrm{mm}\times350~\mathrm{mm}\times10~\mathrm{mm}$.
As shown in Fig.~\ref{fig:scintillator_tiles}, the plates were arranged in a $6 \times 4$ array and optically bonded with optical cement (Eljen EJ-500) to form a $1320~\mathrm{mm} \times 1400~\mathrm{mm} \times 10~\mathrm{mm}$ effectively monolithic scintillator plate.
The effect of the bonded boundaries on the detector performance was evaluated with a prototype and found to be negligible~\cite{FROSTPrototype}.
The scintillator surfaces were coated with TiO$_2$-based reflective paint (Eljen EJ-510) to increase the light collection efficiency.

\subsubsection{Wavelength-shifting fiber}
\label{sec:wlsfiber}

The detector uses Kuraray Y-11(200)M wavelength-shifting fibers with a diameter of $2.0~\mathrm{mm}$ and lengths of $1.35$ and $1.43~\mathrm{m}$ in the horizontal and vertical directions, respectively.
The main properties of the WLS fiber are summarized in Table~\ref{tab:wlsfiber}.
Grooves with a width and depth of $2.5~\mathrm{mm}$ were machined at a pitch of $10~\mathrm{mm}$ on both surfaces of the scintillator plate in orthogonal directions.
The WLS fibers were optically bonded to the scintillator in the grooves using optical cement (Eljen EJ-500).
One end of each fiber was coupled to a SiPM, while the opposite end was aluminized by vapor deposition to reflect light back toward the SiPM and increase the collected light yield.

\begin{table}[htbp]
  \centering
  \caption{Main properties of the Kuraray Y-11(200)M wavelength-shifting fiber used in the NINJA FROST detector~\cite{KurarayPSF}.}
  \label{tab:wlsfiber}
  \begin{tabular}{ll}
    \hline\hline
    Item & Specification \\
    \hline
    Fiber type & Round shape, multi-cladding, S-type \\
    Core material & Polystyrene (PS) \\
    Inner-cladding material & Polymethylmethacrylate (PMMA) \\
    Outer-cladding material & Fluorinated polymer (FP) \\
    Refractive index (core / inner / outer) & 1.59 / 1.49 / 1.42 \\
    Absorption wavelength & $430~\mathrm{nm}$ (peak) \\
    Emission wavelength & $476~\mathrm{nm}$ (peak) \\
    Attenuation length & $>3.5~\mathrm{m}$ \\
    \hline\hline
  \end{tabular}
\end{table}

\subsubsection{Silicon photomultiplier}
\label{sec:sipm}

The scintillation light transported by each WLS fiber is detected by a Hamamatsu S13360-3075CS SiPM.
The main properties of the SiPM are summarized in Table~\ref{tab:sipm}.
The SiPMs were operated at a bias voltage of $55~\mathrm{V}$.

\begin{table}[htbp]
  \centering
  \caption{Main properties of the Hamamatsu S13360-3075CS SiPM~\cite{HamamatsuSiPM}. The characteristics were measured at $25^\circ\mathrm{C}$ with a bias voltage of $V_{\mathrm{BR}}+3~\mathrm{V}$.}
  \label{tab:sipm}
  \begin{tabular}{ll}
    \hline\hline
    Item & Specification \\
    \hline
    Effective photosensitive area & $3.0~\mathrm{mm} \times 3.0~\mathrm{mm}$ \\
    Pixel pitch & $75~\mathrm{\mu m}$ \\
    Number of pixels & 1600 \\
    Fill factor & $82\%$ \\
    Spectral response range & $270$--$900~\mathrm{nm}$ \\
    Peak sensitivity wavelength & $450~\mathrm{nm}$ \\
    Photon detection efficiency & $50\%$ at $450~\mathrm{nm}$ \\
    Gain & $4.0 \times 10^{6}$ \\
    Breakdown voltage ($V_\mathrm{BR}$) & $53 \pm 5~\mathrm{V}$ \\
    Dark count & $500~\mathrm{kcps}$ (typ.) \\
    Crosstalk probability & $7\%$ \\
    \hline\hline
  \end{tabular}
\end{table}

\subsubsection{Mechanical support structure}
\label{sec:support}

\begin{figure}[htbp]
  \centering
  \includegraphics[width=1.0\linewidth]{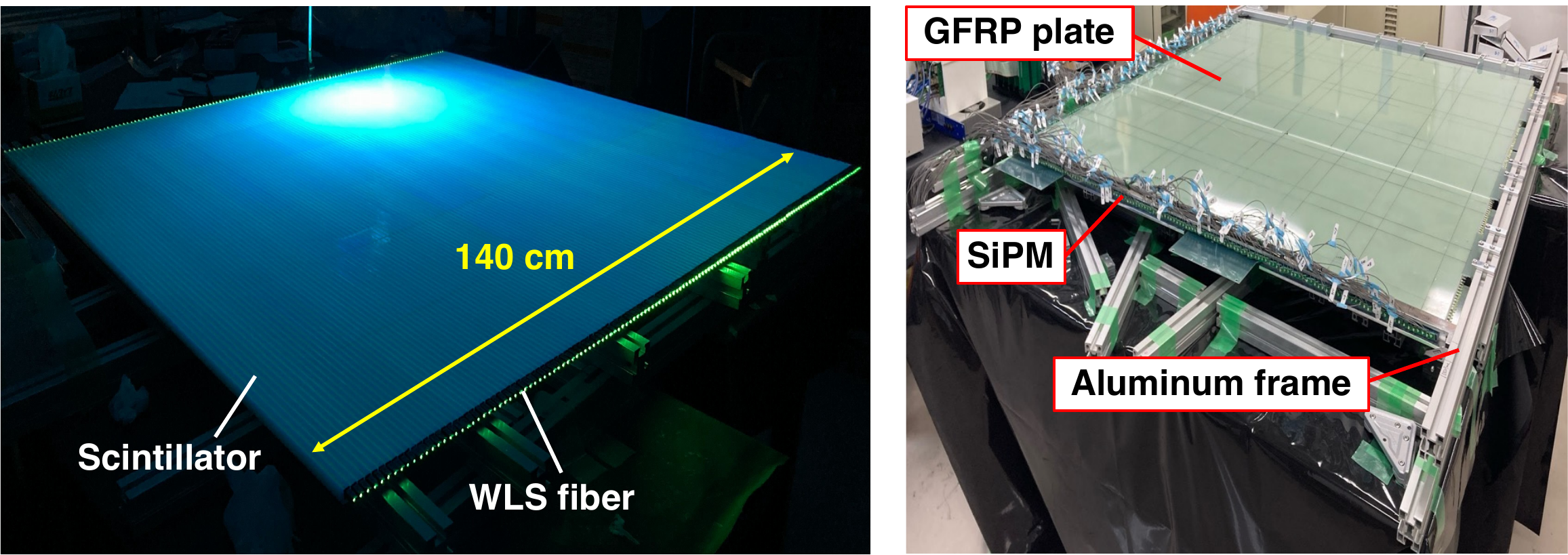}
  \caption{Photographs of the NINJA FROST detector after the WLS fibers were bonded to one surface and illuminated with ultraviolet light (left), and after completion of the detector assembly (right).}
  \label{fig:frost_photos}
\end{figure}

To maintain the mechanical rigidity of the large scintillator plate and suppress deformation, $2~\mathrm{mm}$-thick glass-fiber-reinforced epoxy (GFRP) plates were attached to both sides of the scintillator.
The detector was fixed in an aluminum frame, which was mounted on the NINJA detector rack (see Fig.~\ref{fig:ninja_wagasci_setup}).
Photographs of the NINJA FROST detector during assembly and after completion are shown in Fig.~\ref{fig:frost_photos}.

\subsection{Data acquisition (DAQ) system}
\label{sec:daq}
The signals from the 272 SiPMs were read out using 11 RAYRAW-v1 boards~\cite{RAYRAWManual}.
RAYRAW-v1 is a general-purpose SiPM readout board equipped with four 8-channel YAENAMI ASICs (application-specific integrated circuits) and an FPGA (field-programmable gate array), providing 32 readout channels per board.
The SiPM signal in each channel was amplified and shaped by YAENAMI and digitized as a waveform by a 10-bit ADC (analog-to-digital converter) at a sampling frequency of $75~\mathrm{MHz}$.
The timing of the SiPM signals was recorded by a multi-hit TDC (time-to-digital converter) with a time bin of $0.833~\mathrm{ns}$.

The J-PARC neutrino beam was delivered in spills at intervals of approximately $1.3~\mathrm{s}$, with each spill consisting of eight bunches distributed over $4.1~\mathrm{\mu s}$.
A beam trigger signal provided for each spill was used as an external trigger for the DAQ system.
For each trigger, the ADC waveforms and TDC information were recorded within a $5.3~\mathrm{\mu s}$ acquisition window covering the entire beam spill.

The waveform charge was obtained by integrating the ADC samples for each beam bunch.
The pedestal and single-photoelectron gain were determined for each channel from SiPM dark-noise data, and the integrated waveform charge was converted to the number of photoelectrons (p.e.).

\subsection{Channel response calibration}

To reduce channel-to-channel variations arising from individual WLS fibers and SiPMs, the response of each channel was measured with cosmic rays before the fibers were bonded to the scintillator plate, following the procedure used in the prototype study~\cite{FROSTPrototype}.
A plastic scintillator bar with dimensions of $50~\mathrm{mm} \times 1200~\mathrm{mm} \times 10~\mathrm{mm}$ and holes approximately $2.2~\mathrm{mm}$ in diameter was used for the measurement.
For each channel, the WLS fiber, with the SiPM coupled to one end, was inserted into a hole in the scintillator bar, and the cosmic-ray light yield was measured.
The light yield averaged over all channels was $80.0~\mathrm{p.e.}$, with a relative standard deviation of $8.38\%$.
The measured channel responses were used to determine correction factors that normalized the light yield of each channel to the average response, and the resulting corrected light yield was used in the subsequent analysis.

\section{Optical simulation}
\label{sec:simulation}

\subsection{Simulation framework}
\label{sec:optical_simulation}

A Geant4-based~\cite{geant4:2003,geant4:2006,geant4:2016} Monte Carlo (MC) simulation was used to model the optical response of the NINJA FROST detector.
The simulation framework developed in the previous FROST study~\cite{FROSTPrototype} was extended to reproduce the meter-scale detector geometry described in Section~\ref{sec:designandcomponents}.
To reduce the computational cost, the optically bonded boundaries between the 24 scintillator plates were not modeled, and the scintillator was treated as a single continuous volume.
This approximation was justified by the prototype study, in which the effect of the bonded boundaries on the detector performance was found to be negligible~\cite{FROSTPrototype}.

Optical photons were individually tracked in Geant4 throughout the detector geometry.
The simulation included photon scattering in the scintillator, reflections at its surfaces, absorption and re-emission in the WLS fibers, and photon detection by the SiPMs.
The effects of SiPM dark counts, optical crosstalk, afterpulsing, and saturation due to the finite number of pixels were also included.
The embedded scatterers were modeled as Rayleigh scattering characterized by an effective scattering length, $\lambda_{\mathrm{scat}}$.
Light attenuation along the WLS fibers was modeled as photon absorption in the fibers and characterized by an attenuation length, $\lambda_{\mathrm{att}}$.
The scintillation photon yield was characterized by $Y_{\mathrm{scint}}$, defined as the number of photons produced per unit deposited energy.

Most optical parameters were fixed to values obtained from manufacturer datasheets.
Three parameters that strongly affect the light-yield distribution were tuned using cosmic-ray data: the effective scattering length $\lambda_{\mathrm{scat}}$, the attenuation length of the WLS fiber $\lambda_{\mathrm{att}}$, and the scintillation photon yield $Y_{\mathrm{scint}}$.
The cosmic-ray data acquisition setup used for the optical-parameter tuning is described in Section~\ref{sec:tuning_measurement}, and the tuning procedure is presented in Section~\ref{sec:parameter_tuning}.

\subsection{Cosmic-ray data acquisition setup for optical-parameter tuning}
\label{sec:tuning_measurement}

\begin{figure}[htbp]
  \centering
  \includegraphics[width=1.0\linewidth]{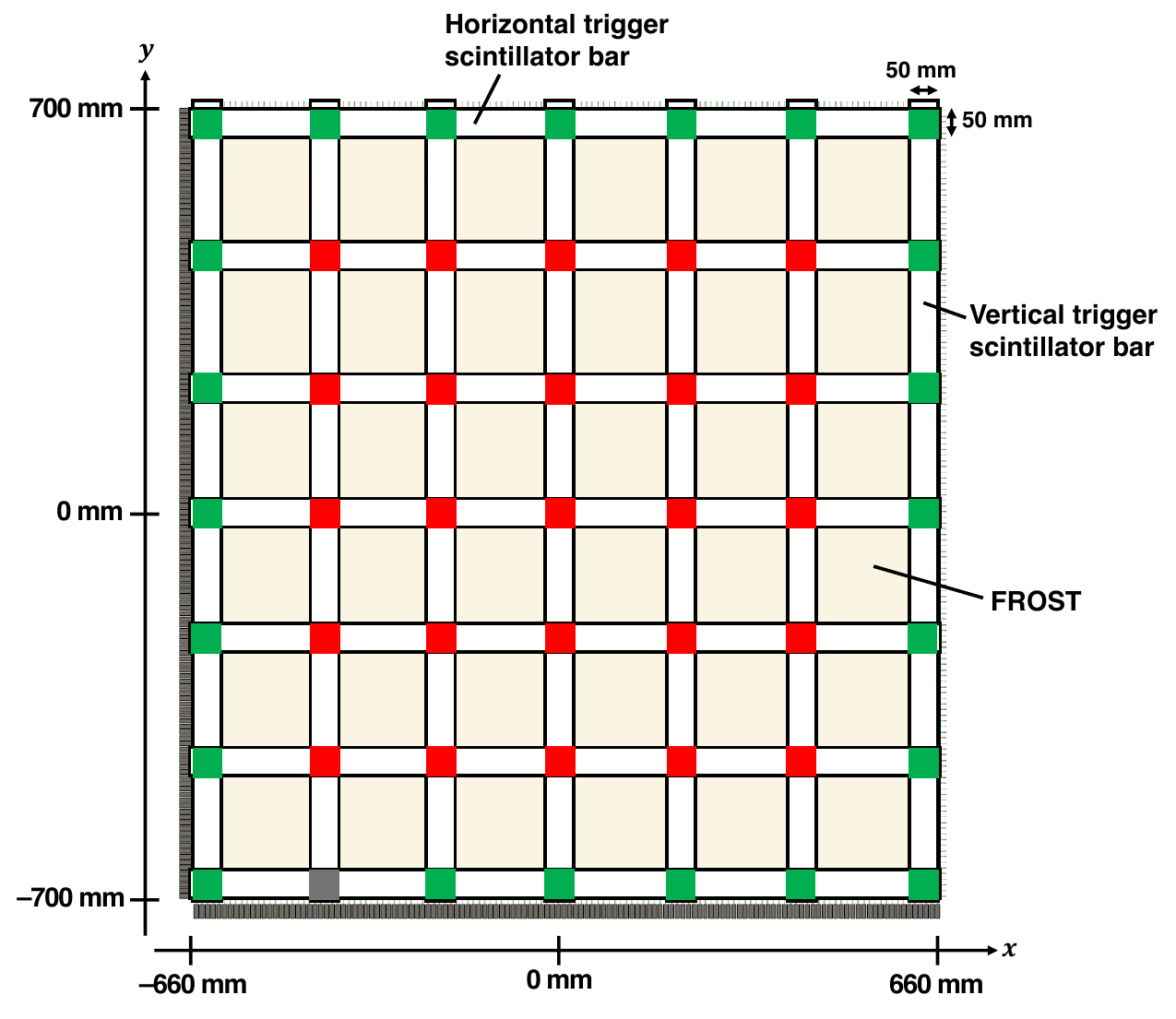}
  \caption{Schematic setup of the cosmic-ray data acquisition used for optical-simulation tuning. The red regions were used to tune the scattering length. The red and green regions were used to tune the attenuation length of the WLS fiber and the scintillation photon yield. The gray region was excluded because of insufficient statistics.}
  \label{fig:tuning_setup}
\end{figure}

Cosmic-ray data for the parameter tuning were taken with the NINJA FROST detector placed horizontally and two orthogonal layers of plastic scintillator bars installed above it.
Each layer consisted of seven scintillator bars with a width of $50~\mathrm{mm}$ and a thickness of $10~\mathrm{mm}$.
The signals from the seven bars in each layer were combined with a logical OR, and the coincidence of the OR signals from the two layers was used as the trigger.
For each event, the muon crossing region was identified when exactly one trigger scintillator in each layer had a light yield above $20~\mathrm{p.e.}$
The intersection of the two selected scintillator bars defined the muon crossing region on the scintillator plate.
The two sets of seven scintillator bars therefore defined 49 crossing regions, each with dimensions of $50~\mathrm{mm}\times50~\mathrm{mm}$, on the scintillator plate.
For the cosmic-ray sample used in the parameter tuning, events were additionally required to have a maximum light yield exceeding $20~\mathrm{p.e.}$ in each of the $x$ and $y$ readout directions of the NINJA FROST detector.
The data acquisition setup and the definition of the crossing regions are shown in Fig.~\ref{fig:tuning_setup}.
In the MC, $1.0~\mathrm{GeV}$ muons were generated uniformly in position over the scintillator plate with a zenith-angle distribution proportional to $\cos^{2}\theta$, where $\theta$ is the zenith angle~\cite{PDG2022}.

\subsection{Tuning of optical parameters using cosmic-ray data}
\label{sec:parameter_tuning}

Two observables were used to tune the optical parameters: the width of the light-yield distribution across the readout channels and the maximum light yield among the readout channels.
The width of the light-yield distribution was used to tune $\lambda_{\mathrm{scat}}$, while the maximum light yield was used to tune $\lambda_{\mathrm{att}}$ and $Y_{\mathrm{scint}}$.

\subsubsection{Effective scattering length $\lambda_\mathrm{scat}$}

The light-yield distribution in each event was fitted with a Gaussian function, and the fitted width $\sigma_{\mathrm{LY}}$ was used to characterize the spatial spread of the collected light.
For each crossing region and readout direction, the mean value of $\sigma_{\mathrm{LY}}$ was calculated using events with $\sigma_\mathrm{LY} < 15~\mathrm{mm}$ to suppress the influence of outliers arising from failed Gaussian fits.

The effective scattering length $\lambda_{\mathrm{scat}}$ was tuned using the 25 central crossing regions indicated in red in Fig.~\ref{fig:tuning_setup}.
Regions near the detector edges were excluded to minimize edge effects on the light-yield distribution.
For each crossing region and readout direction, the mean of the event-by-event fitted widths $\sigma_{\mathrm{LY}}$ was calculated from MC samples generated at several values of $\lambda_{\mathrm{scat}}$.
The value of $\lambda_{\mathrm{scat}}$ reproducing the mean $\sigma_{\mathrm{LY}}$ measured in data was determined by linear interpolation of the MC results.
The distribution of the resulting $\lambda_{\mathrm{scat}}$ values from the 25 crossing regions and both readout directions was fitted with a Gaussian function.
The mean of the Gaussian fit was taken as the tuned scattering length.

\subsubsection{WLS-fiber attenuation length $\lambda_\mathrm{att}$}

The attenuation length $\lambda_{\mathrm{att}}$ was tuned using the position dependence of the maximum light yield.
All crossing regions shown in red and green in Fig.~\ref{fig:tuning_setup} were used, while the gray region was excluded because of insufficient statistics.
For each crossing region and readout direction, the distribution of the maximum light yield was fitted with a Landau function, and its most probable value (MPV) was used for the tuning.

For each trial value of $\lambda_{\mathrm{att}}$, the following loss function was evaluated:
\begin{equation}
    L_{\mathrm{att}}
    =
    \sum_i
    \frac{\left(D_i-aM_i\right)^2}{D_i}.
\end{equation}
Here, $D_i$ and $M_i$ are the MPVs of the maximum-light-yield distributions in data and MC, respectively, and the index $i$ runs over the crossing regions and readout directions.
The normalization factor $a$ was chosen to minimize $L_{\mathrm{att}}$ for each trial value, so that $\lambda_{\mathrm{att}}$ was determined primarily from the position dependence of the light yield rather than its overall normalization.
The value of $\lambda_{\mathrm{att}}$ giving the minimum loss was selected.

\subsubsection{Scintillation photon yield $Y_\mathrm{scint}$}

After fixing $\lambda_{\mathrm{att}}$, the scintillation photon yield $Y_{\mathrm{scint}}$ was tuned using the absolute scale of the maximum-light-yield MPVs.
The same crossing regions as those used for the tuning of $\lambda_{\mathrm{att}}$ were used.
For each trial value of $Y_{\mathrm{scint}}$, the loss function
\begin{equation}
    L_{\mathrm{scint}}
    =
    \sum_i
    \frac{\left(D_i-M_i\right)^2}{D_i}
\end{equation}
was evaluated.
Unlike the tuning of $\lambda_{\mathrm{att}}$, no additional normalization factor was introduced, since the difference in the absolute light-yield scale between data and MC was used to constrain $Y_{\mathrm{scint}}$.
The value of $Y_{\mathrm{scint}}$ giving the minimum loss was selected.

Because $\lambda_{\mathrm{scat}}$, $\lambda_{\mathrm{att}}$, and $Y_{\mathrm{scint}}$ are not independent, the three parameters were tuned sequentially and the procedure was repeated until convergence.
The tuning results were confirmed to have converged after the second iteration, and the corresponding parameter values were adopted as the final tuned values, as summarized in Table~\ref{tab:optical_tuning}.

\begin{table}[htbp]
  \centering
  \caption{
  Final optical parameters obtained from the tuning.
  }
  \label{tab:optical_tuning}
  \begin{tabular}{ll}
    \hline\hline
    Parameter & Tuned value \\
    \hline
    Effective scattering length, $\lambda_{\mathrm{scat}}$
      & $0.53~\mathrm{mm}$ \\
    Attenuation length of the WLS fiber, $\lambda_{\mathrm{att}}$
      & $3.8~\mathrm{m}$ \\
    Scintillation photon yield, $Y_{\mathrm{scint}}$
      & $7400~\mathrm{photons/MeV}$ \\
    \hline\hline
  \end{tabular}
\end{table}

\subsubsection{Tuning results}

\begin{figure}[htbp]
  \centering
  \includegraphics[width=0.6\linewidth]{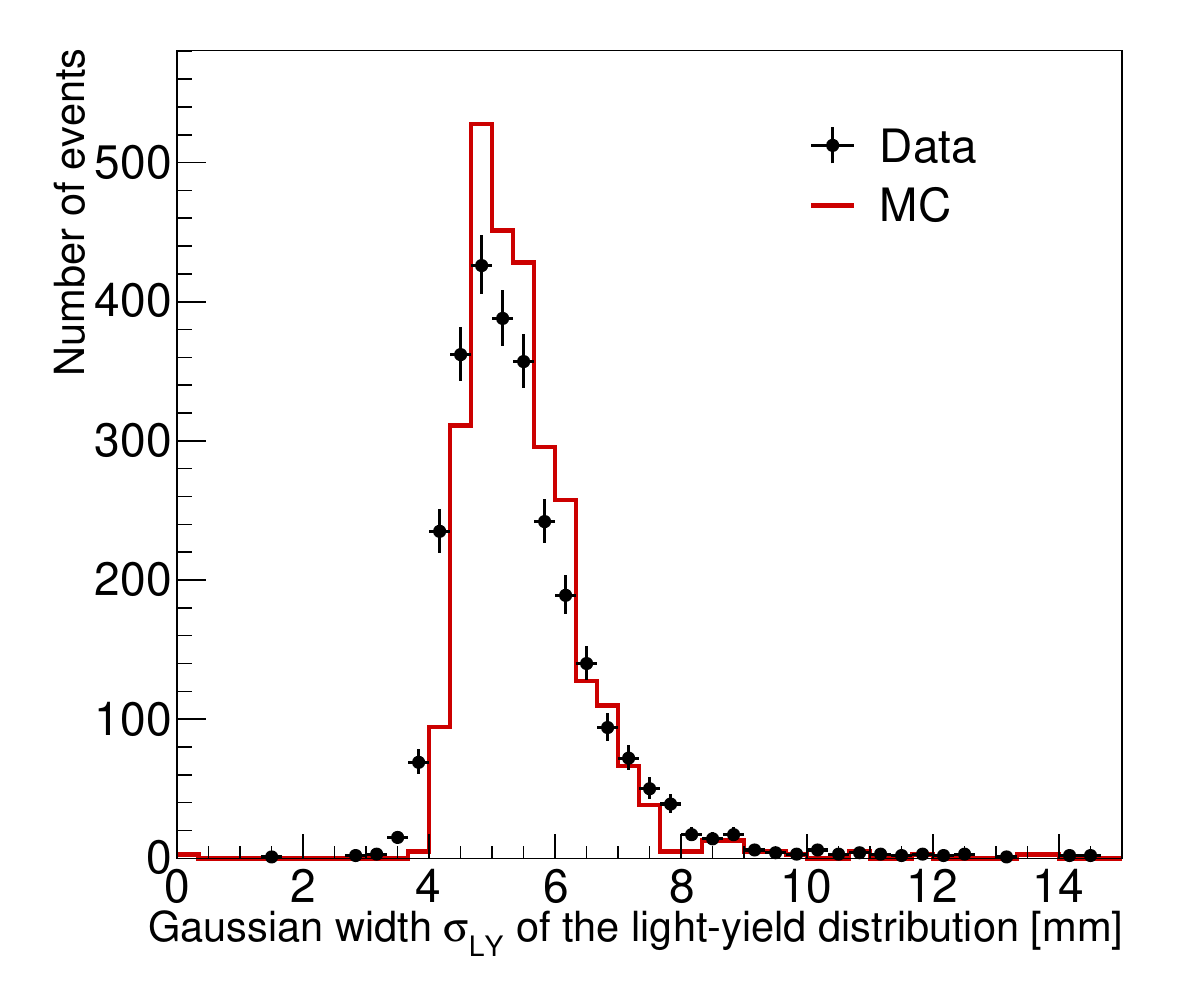}
  \caption{Comparison of the $\sigma_{\mathrm{LY}}$ distributions between data and the tuned MC for the $x$ readout direction in the central crossing region, $-30~\mathrm{mm} \leq x \leq 20~\mathrm{mm}$ and $-30~\mathrm{mm} \leq y \leq 20~\mathrm{mm}$. The MC distribution is normalized to the total number of events in the data distribution.}
  \label{fig:tuned_sigma}
\end{figure}

Figure~\ref{fig:tuned_sigma} compares the $\sigma_{\mathrm{LY}}$ distributions in data and MC with the tuned parameters for the $x$ readout direction in the central crossing region, $-30~\mathrm{mm} \leq x \leq 20~\mathrm{mm}$ and $-30~\mathrm{mm} \leq y \leq 20~\mathrm{mm}$.
The tuned MC provides a reasonable description of the measured $\sigma_{\mathrm{LY}}$ distribution in this representative central region.

\begin{figure}[htbp]
  \centering
  \includegraphics[width=1.0\linewidth]{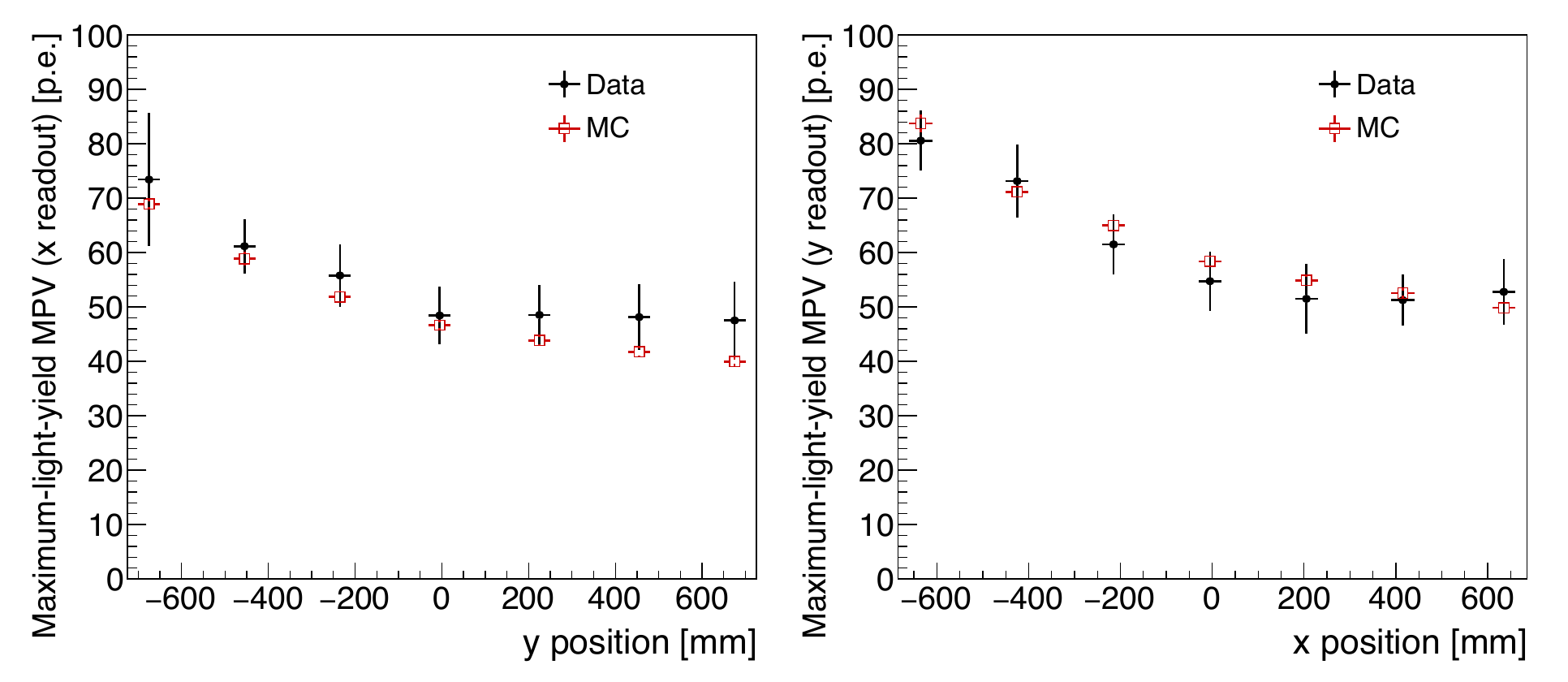}
  \caption{Position dependence of the maximum-light-yield MPV in data and the tuned MC. The left and right panels show the $x$- and $y$-readout results, respectively. The vertical error bars indicate the standard deviation among the crossing regions at each position, and the horizontal error bars indicate the extent of the crossing regions.}
  \label{fig:tuned_maxlight}
\end{figure}

The position dependence of the maximum light yield was also compared between data and the tuned MC.
The fibers used for the $x$ readout run along the $y$ direction, while those used for the $y$ readout run along the $x$ direction.
Therefore, the maximum-light-yield MPV for the $x$ readout is shown as a function of the $y$ position, and that for the $y$ readout is shown as a function of the $x$ position in Fig.~\ref{fig:tuned_maxlight}.
For each position, the MPVs obtained in the crossing regions at different positions in the orthogonal direction were averaged.
The vertical error bars indicate the standard deviation of the MPVs among these crossing regions, while the horizontal error bars indicate the $\pm25~\mathrm{mm}$ extent of the crossing regions.
The tuned MC reproduces the overall position dependence of the maximum light yield observed in data for both readout directions.

\section{Position reconstruction}
\label{sec:reconstruction}

\subsection{Position reconstruction method}
\label{sec:position_reconstruction}

\begin{figure}[htbp]
  \centering
  \includegraphics[width=1.0\linewidth]{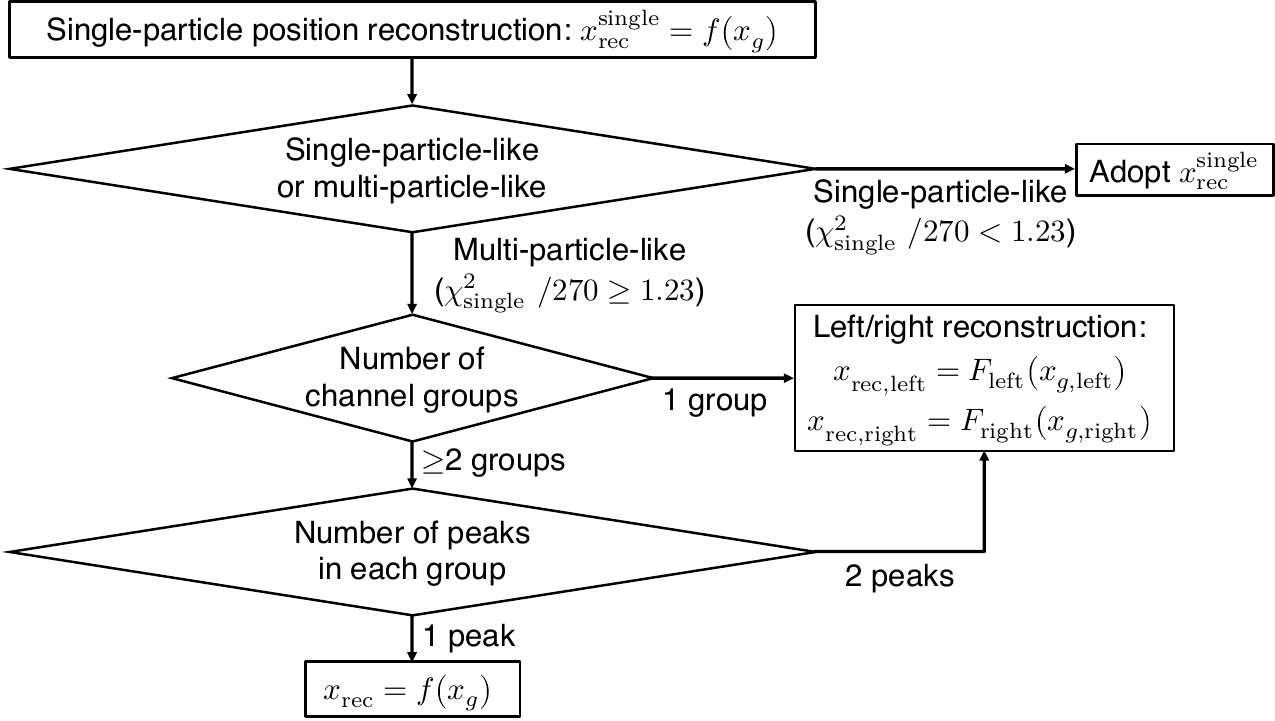}
  \caption{Flowchart of the position reconstruction procedure.}
\label{fig:position_reconstruction_flowchart}
\end{figure}

The position reconstruction was designed to handle both single- and multi-particle events.
The single-particle reconstruction follows the basic method developed in the previous FROST study~\cite{FROSTPrototype}, while a dedicated multi-particle reconstruction was newly developed in this work.
In the neutrino-interaction MC described in Section~\ref{sec:reconstruction_simulation}, $67.4\%$ of the selected simulated neutrino-interaction events in the ECCs had two or more particles crossing the NINJA FROST detector, making a reconstruction method for multi-particle events essential.
For each event, the position was first reconstructed under the single-particle hypothesis to obtain the expected light-yield distribution used for the subsequent single- and multi-particle classification.
The event was then classified as single- or multi-particle-like by comparing the observed light-yield distribution with that expected under the single-particle hypothesis.
For single-particle-like events, the position obtained under the single-particle hypothesis was adopted, while multi-particle-like events were processed with the dedicated multi-particle reconstruction.
The three steps are described below and summarized in Fig.~\ref{fig:position_reconstruction_flowchart}.

\subsubsection{Single-particle position reconstruction}
\label{sec:single_reconstruction}

The single-particle position was reconstructed from the light-yield distribution across the readout channels using the weighted-center method.
For the $x$ direction, a weighted center of light yield, $x_g$, was calculated as
\begin{equation}
    x_g
    =
    \frac{\displaystyle\sum_i (N_i^{\mathrm{obs}})^{\alpha} x_i}
         {\displaystyle\sum_i (N_i^{\mathrm{obs}})^{\alpha}},
    \label{eq:weighted_center}
\end{equation}
where $N_i^{\mathrm{obs}}$ is the observed light yield in channel $i$ and $x_i$ is the position of the corresponding WLS fiber.
Increasing $\alpha$ reduces the relative contribution of channels with small light yields and thereby reduces the effects of statistical fluctuations and noise in these channels on $x_g$.
However, if $\alpha$ is too large, $x_g$ becomes increasingly dominated by the channel with the largest light yield, reducing its sensitivity to the particle crossing position between neighboring fibers.
In the previous FROST study, $\alpha=1$ was used without optimization~\cite{FROSTPrototype}, whereas the larger number of readout channels in the meter-scale detector makes suppressing contributions from channels with small light yields more important.
In this study, $\alpha$ was optimized using simulation as described in Section~\ref{sec:reconstruction_simulation}, and $\alpha=4.5$ was adopted.

The weighted center of light yield $x_g$ does not directly correspond to the particle crossing position.
A mapping function from the average weighted center of light yield $\overline{x_g}$ to the true crossing position $x_{\mathrm{true}}$ was therefore determined using the optical simulation described in Section~\ref{sec:simulation}:
\begin{equation}
    x_{\mathrm{true}} = f(\overline{x_g}).
    \label{eq:mapping_function}
\end{equation}

\begin{figure}[htbp]
  \centering
  \includegraphics[width=0.6\linewidth]{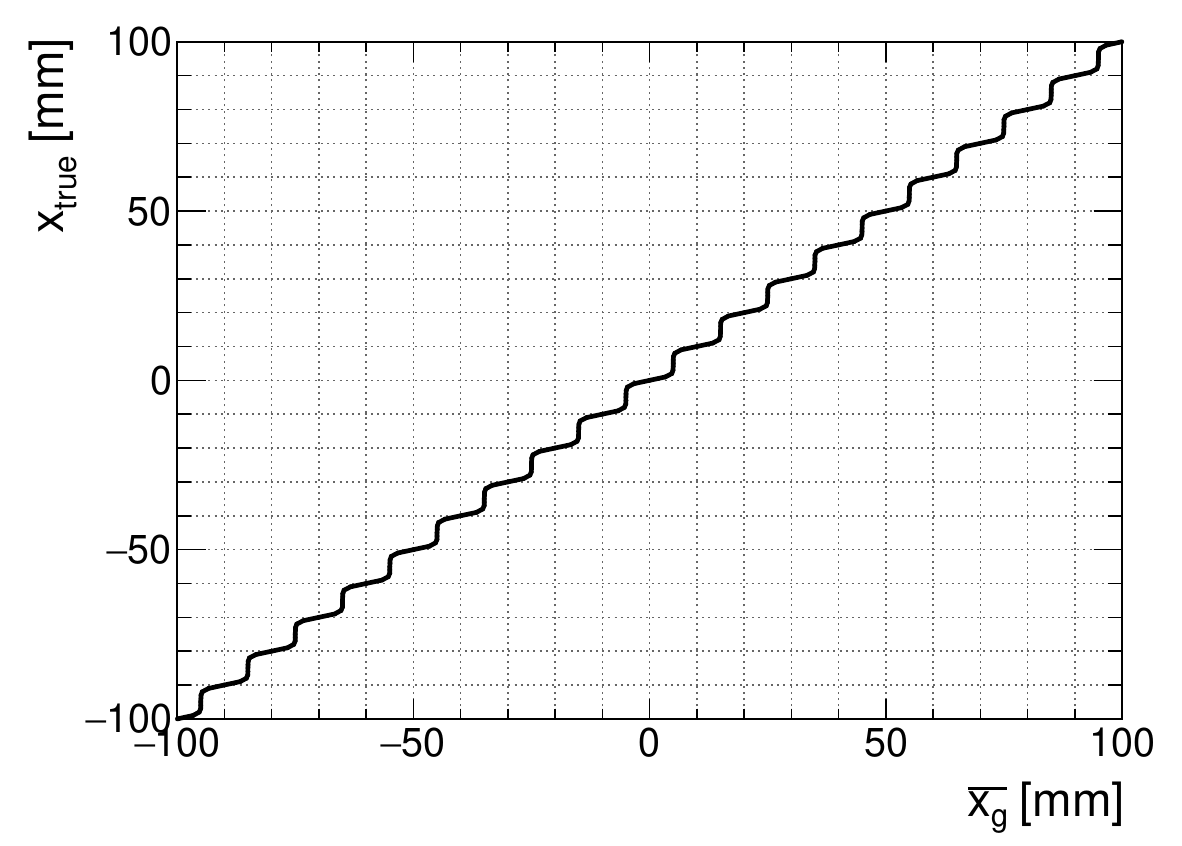}
  \caption{Mapping function $x_{\mathrm{true}} = f(\overline{x_g})$ in the range $-100~\mathrm{mm}<\overline{x_g}<100~\mathrm{mm}$.}
  \label{fig:mapping_function}
\end{figure}

To construct the mapping function, $1.0~\mathrm{GeV}$ muons were injected perpendicular to the scintillator at selected $x_{\mathrm{true}}$ positions, with $y_{\mathrm{true}}$ uniformly distributed within $-5~\mathrm{mm}<y_{\mathrm{true}}<5~\mathrm{mm}$.
From these simulation results, the relation between $x_{\mathrm{true}}$ and the average weighted center of light yield $\overline{x_g}$ was constructed at $1~\mathrm{mm}$ intervals, and the mapping function $f$ was defined by linear interpolation between the resulting points.
Although the relation showed a weak dependence on $y_{\mathrm{true}}$, a one-dimensional mapping function was used because including this dependence would substantially increase the simulation cost and reconstruction complexity.
The resulting mapping function is shown in Fig.~\ref{fig:mapping_function}.

For each event, the reconstructed position was obtained by applying this mapping function to the measured weighted center of light yield:
\begin{equation}
    x_{\mathrm{rec}}^{\mathrm{single}} = f(x_g).
    \label{eq:position_reconstruction}
\end{equation}
The $y$ position was reconstructed using the same procedure.

\subsubsection{Single- and multi-particle event classification}
\label{sec:particle_classification}

To distinguish single- and multi-particle events, the observed light-yield distribution was compared with the distribution expected under the single-particle hypothesis at the reconstructed position obtained in Section~\ref{sec:single_reconstruction}.
The expected light yield in each channel as a function of the particle crossing position was obtained by averaging the light yield at each simulated crossing position in the single-muon MC samples used to construct the mapping function.
For the $x$ and $y$ readout channels, the expected light yields were evaluated at $x_{\mathrm{rec}}^{\mathrm{single}}$ and $y_{\mathrm{rec}}^{\mathrm{single}}$, respectively.
The observed and expected light yields were then compared using a Poisson-likelihood $\chi^2$ defined as
\begin{equation}
    \chi^2_{\mathrm{single}}
    \left(
        x_{\mathrm{rec}}^{\mathrm{single}},
        y_{\mathrm{rec}}^{\mathrm{single}}
    \right)
    =
    2\sum_i
    \left[
        N_i^{\mathrm{exp}}
        -
        N_i^{\mathrm{obs}}
        +
        N_i^{\mathrm{obs}}
        \ln
        \left(
            \frac{N_i^{\mathrm{obs}}}
                 {N_i^{\mathrm{exp}}}
        \right)
    \right],
    \label{eq:single_chi2}
\end{equation}
where $N_i^{\mathrm{obs}}$ and $N_i^{\mathrm{exp}}$ are the observed and expected light yields, respectively, in channel $i$.
The sum was taken over all 132 and 140 channels in the $x$ and $y$ readout directions, respectively.
Accounting for the two reconstructed position parameters, the number of degrees of freedom was taken as $132+140-2=270$.

For single-particle events, the observed light-yield distribution is expected to be well described by the single-particle hypothesis, resulting in a relatively small value of
$\chi^2_{\mathrm{single}}(x_{\mathrm{rec}}^{\mathrm{single}},y_{\mathrm{rec}}^{\mathrm{single}})/270$.
In contrast, multi-particle events generally give larger values because the superposition of light from multiple particles differs from the distribution expected under the single-particle hypothesis.

Misclassifying a multi-particle event as single-particle-like is more problematic than the opposite case, because the single-particle reconstruction can yield a position between the particle crossing positions, whereas the multi-particle reconstruction can reconstruct a single-particle event with only a limited degradation in position resolution.
In the simulation study described in Section~\ref{sec:reconstruction_simulation}, the classification threshold was therefore determined using simulated single-particle events and two-particle events whose crossing positions in the detector were separated by at least $11.2~\mathrm{mm}$ along either the $x$ or $y$ direction, corresponding to twice the required position resolution of $5.6~\mathrm{mm}$.

\begin{figure}[htbp]
  \centering
  \includegraphics[width=0.6\linewidth]{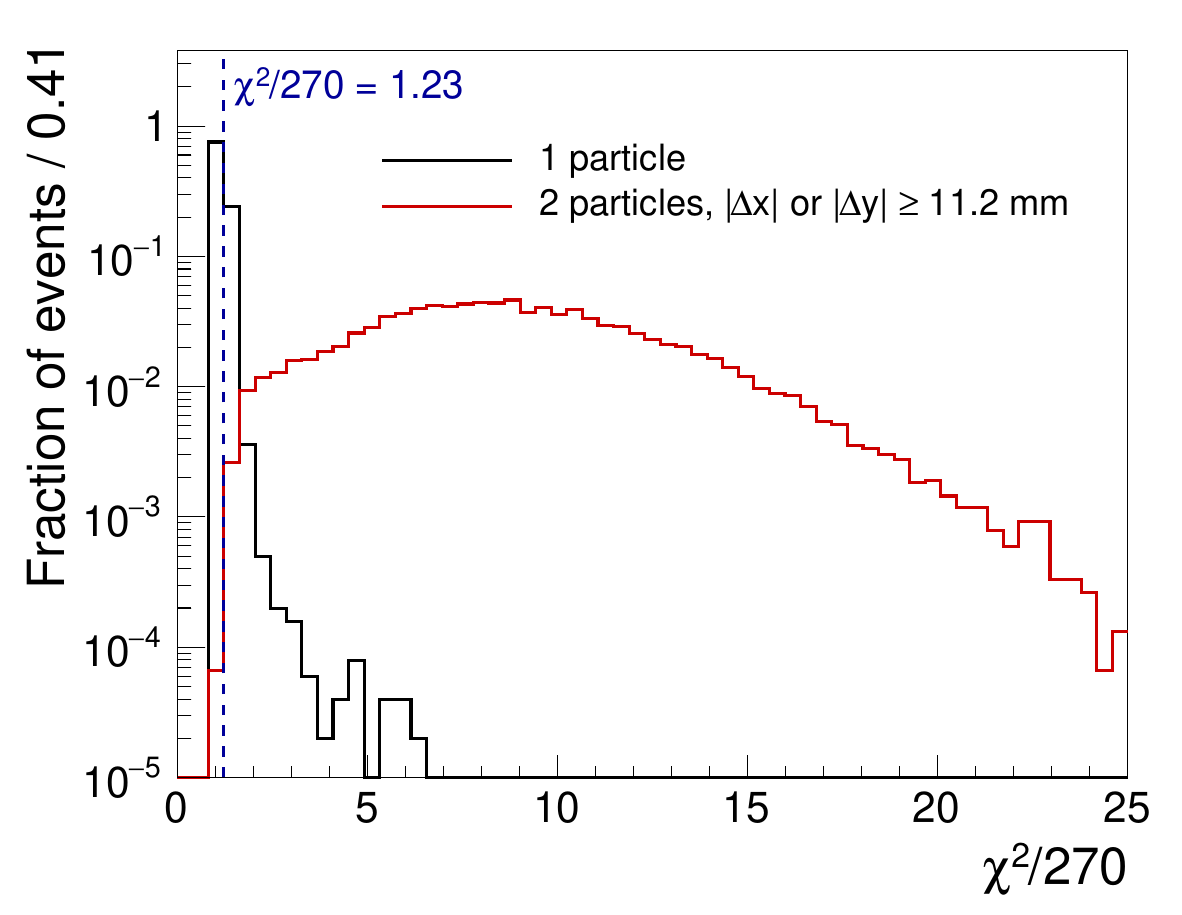}
  \caption{Distributions of $\chi^2_{\mathrm{single}}(x_{\mathrm{rec}}^{\mathrm{single}},y_{\mathrm{rec}}^{\mathrm{single}})/270$ for simulated single-particle events and two-particle events with a separation of at least $11.2~\mathrm{mm}$ between the particle crossing positions in the detector along either the $x$ or $y$ direction. The dashed vertical line indicates the classification threshold at $\chi^2/270=1.23$.
  }
  \label{fig:single_multi_chi2}
\end{figure}

Figure~\ref{fig:single_multi_chi2} shows the distributions of
$\chi^2_{\mathrm{single}}(x_{\mathrm{rec}}^{\mathrm{single}},y_{\mathrm{rec}}^{\mathrm{single}})/270$
for the two simulated event samples.
The classification threshold was set to 1.23 so that the fraction of two-particle events satisfying the $11.2~\mathrm{mm}$ separation criterion that were misclassified as single-particle-like was $0.01\%$.
Events below this threshold were classified as single-particle-like, while the remaining events were classified as multi-particle-like.
In the simulation study described in Section~\ref{sec:reconstruction_simulation}, this criterion correctly classified $75.5\%$ of single-particle events as single-particle-like.
For events classified as single-particle-like, $(x_{\mathrm{rec}}^{\mathrm{single}},y_{\mathrm{rec}}^{\mathrm{single}})$ was adopted as the reconstructed muon position in the detector.

\subsubsection{Multi-particle position reconstruction}
\label{sec:multi_reconstruction}

For events classified as multi-particle-like, multiple position candidates were reconstructed independently in the $x$ and $y$ directions.
The reconstruction consists of three steps: identifying peaks in the light-yield distribution, calculating a weighted center of light yield for each position candidate, and converting the weighted centers to reconstructed positions.
The procedure is described below for the $x$ direction.
The same procedure was applied to the $y$ direction.
The numerical values used in the following reconstruction criteria were chosen empirically based on simulation studies.

First, channels with a light yield of at least $10~\mathrm{p.e.}$ were selected to suppress noise while retaining the particle-induced signals and were grouped into sets of consecutive channel indices.
When particles cross the detector at nearby positions, their light-yield distributions can overlap, causing signals from multiple particles to be included in a single group.
Each group was therefore further examined using up to the six channels with the largest light yields in that group.
Defining $N_{\max}$ as the maximum light yield in the group, the threshold for separating nearby peaks was defined as
\begin{equation}
  N_{\mathrm{th}}=\min\left(\frac{N_{\max}}{4}, 20~\mathrm{p.e.}\right).
  \label{eq:multi_peak_threshold}
\end{equation}
The $N_{\max}/4$ term scales the threshold with the group light yield, allowing the criterion to be applied to groups with different overall light yields.
The threshold was capped at $20~\mathrm{p.e.}$ to retain smaller neighboring peaks when $N_{\max}$ was large.
Among the selected channels, those with
$N_i^{\mathrm{obs}} \geq N_{\mathrm{th}}$
were divided into up to two groups of consecutive channel indices.

\begin{figure}[htbp]
  \centering
  \includegraphics[width=1.0\linewidth]{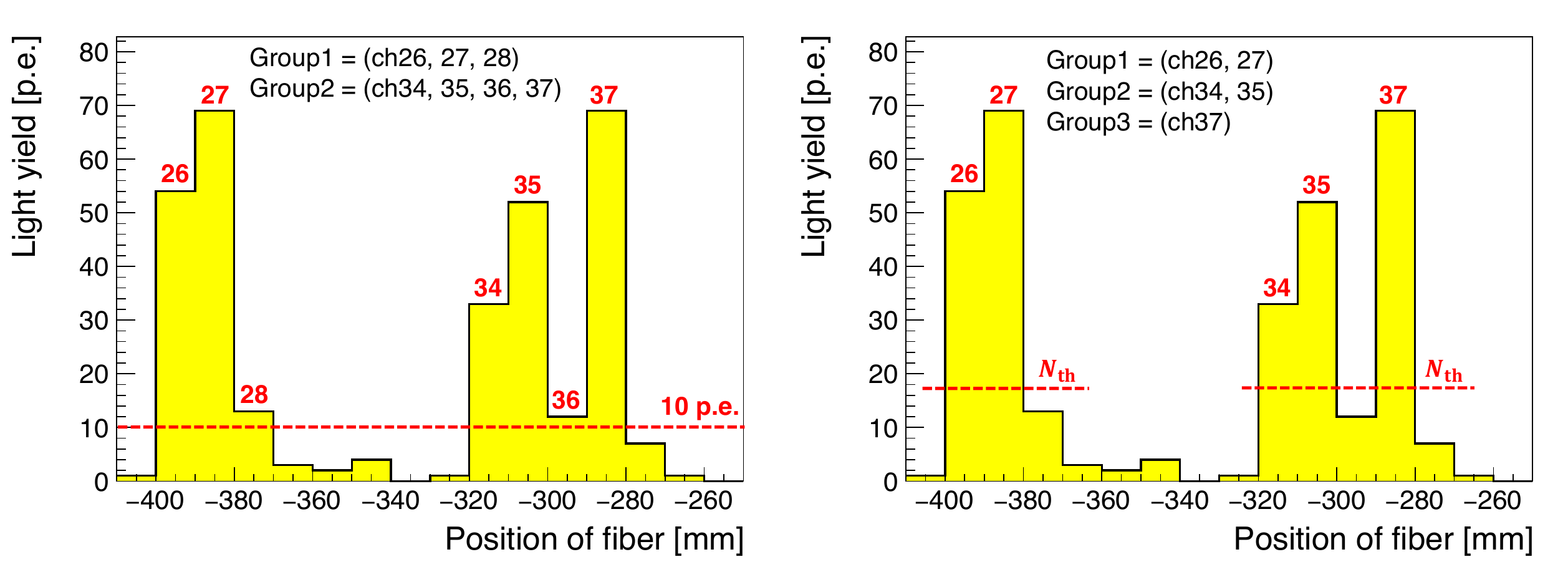}
  \caption{Example of the channel grouping procedure for a simulated multi-particle-like event. The horizontal axis represents the fiber position, and the vertical axis represents the light yield in each readout channel. The numbers above the bins indicate the channel indices. The initial groups obtained with the $10~\mathrm{p.e.}$ threshold are shown on the left, and the groups after applying $N_{\mathrm{th}}$ are shown on the right.}
  \label{fig:multi_reconstruction_example}
\end{figure}

Figure~\ref{fig:multi_reconstruction_example} illustrates the grouping procedure for a representative multi-particle-like event.
As shown in the left panel, the $10~\mathrm{p.e.}$ threshold initially identifies two groups of consecutive channel indices.
One of these groups contains two light-yield peaks that remain connected because the channel between the peaks also exceeds $10~\mathrm{p.e.}$
Applying the additional threshold $N_{\mathrm{th}}$ excludes the low-light-yield channel between the two peaks and separates this group into two, resulting in three groups in total, as shown in the right panel.

After the channel groups were identified, the subsequent procedure depended on the number of identified groups.
If only one group was identified, it was treated as potentially containing unresolved signals from multiple particles because the event had already been classified as multi-particle-like.
In this case, the channel range of the group was divided into two halves, and the weighted centers of light yield $x_{g,\mathrm{left}}$ and $x_{g,\mathrm{right}}$ were calculated for the left and right halves, respectively.

\begin{figure}[htbp]
  \centering
  \includegraphics[width=0.6\linewidth]{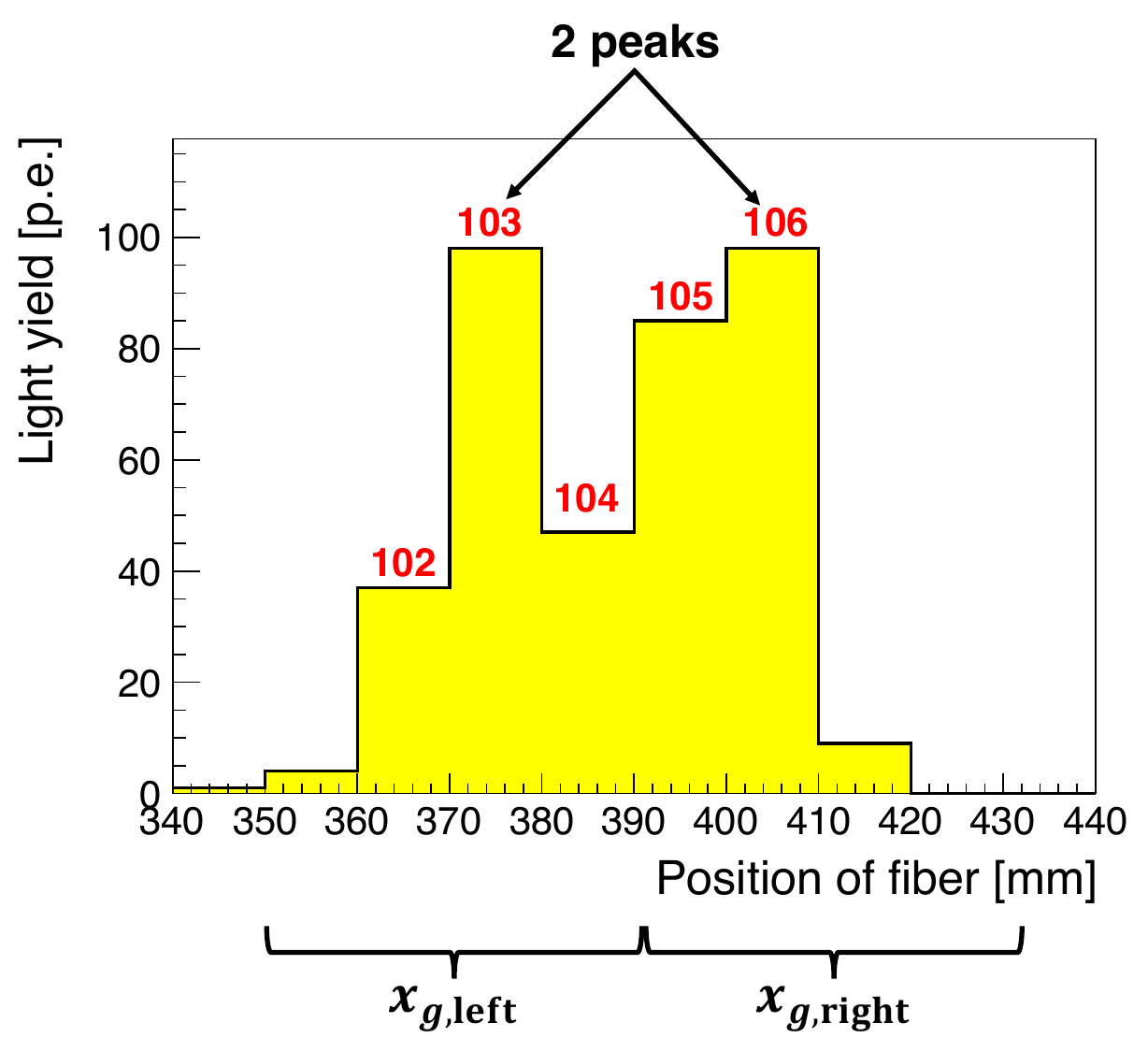}
  \caption{Example of two-peak identification and calculation of the weighted centers of light yield within a channel group in a simulated event. The horizontal axis represents the fiber position, and the vertical axis represents the light yield in each readout channel. The numbers above the bins indicate the channel indices.}
  \label{fig:twopeak_example}
\end{figure}

When two or more groups were identified, each group was examined to determine whether it contained one or two peaks.
Let $c_1$, $c_2$, and $c_3$ denote the channel indices of the three channels with the largest light yields in the group, in descending order of light yield.
A group was regarded as containing two peaks when
\begin{equation}
    |c_1-c_2| \geq 2
    \quad \mathrm{or} \quad
    \left(
        |c_1-c_3| \geq 2
        \ \mathrm{and}\
        |c_2-c_3| \geq 2
    \right).
    \label{eq:two_peak_criterion}
\end{equation}
The first condition identifies two peaks when the two channels with the largest light yields are not adjacent.
If these two channels are adjacent and are therefore regarded as belonging to the same peak, the second condition identifies another peak when the channel with the third-largest light yield is not adjacent to either of them.
If neither condition was satisfied, the group was regarded as containing a single peak.
For a group identified as containing two peaks, the weighted centers of light yield on the left and right sides of the group, $x_{g,\mathrm{left}}$ and $x_{g,\mathrm{right}}$, respectively, were calculated in the same manner as in the case where only one group was identified in the entire light-yield distribution.
For a group identified as containing a single peak, a single weighted center of light yield, $x_g$, was calculated.
Figure~\ref{fig:twopeak_example} shows an example of a group identified as containing two peaks and the corresponding weighted centers of light yield.

When one weighted center $x_g$ was obtained from a group, its reconstructed position was calculated using the single-particle mapping function:
\begin{equation}
    x_{\mathrm{rec}} = f(x_g).
    \label{eq:multi_single_mapping}
\end{equation}

When two weighted centers $x_{g,\mathrm{left}}$ and $x_{g,\mathrm{right}}$ were obtained from the same group, dedicated mapping functions were used because the overlapping light distributions of the two particles modify the relation between the weighted center and the particle crossing position.
The functions $F_\mathrm{left}$ and $F_\mathrm{right}$ were constructed using simulations with two $1.0~\mathrm{GeV}$ muons injected perpendicular to the scintillator.
To construct $F_\mathrm{left}$, $x_{\mathrm{true,left}}$ was varied, while the second muon's $x$ position was sampled uniformly between $x_{\mathrm{true,left}}$ and the right edge of the scintillator plate.
For both muons, $y_{\mathrm{true}}$ was uniformly distributed within $-5~\mathrm{mm}<y_{\mathrm{true}}<5~\mathrm{mm}$.
Events in which the two particle signals formed a single group were selected.
The average $x_{g,\mathrm{left}}$ at each true position was used to construct the relation
\begin{equation}
    x_{\mathrm{true,left}}
    =
    F_\mathrm{left}(\overline{x_{g,\mathrm{left}}}).
\end{equation}
The function $F_\mathrm{right}$ was constructed analogously for the right muon:
\begin{equation}
    x_{\mathrm{true,right}}
    =
    F_\mathrm{right}(\overline{x_{g,\mathrm{right}}}).
\end{equation}
The reconstructed positions were then obtained as
\begin{equation}
    x_{\mathrm{rec,left}} = F_\mathrm{left}(x_{g,\mathrm{left}}),
    \qquad
    x_{\mathrm{rec,right}} = F_\mathrm{right}(x_{g,\mathrm{right}}).
    \label{eq:multi_position_mapping}
\end{equation}

This procedure yields two or more reconstructed position candidates for a multi-particle-like event.
Because the $x$ and $y$ candidates are reconstructed independently, the two-dimensional track positions are not uniquely determined owing to the multiple possible pairings of the reconstructed $x$ and $y$ candidates.
The reconstructed $x$ and $y$ position candidates were retained independently because the ambiguity can be resolved in subsequent track matching with other detectors.

\subsection{Simulation study of the position reconstruction}
\label{sec:reconstruction_simulation}

The position reconstruction method was evaluated using a neutrino interaction simulation.
The neutrino beam flux was simulated with JNUBEAM~\cite{JNUBEAM}, and neutrino interactions in the ECCs were generated with NEUT~\cite{NEUT} using the same settings as those used in the NINJA physics analysis~\cite{NINJAPhysicsRun}.
The detector response was simulated with Geant4 using the WAGASCI--BabyMIND detector simulation framework~\cite{WAGASCIBabyMIND}.
The NINJA detector geometry was implemented in this framework, including the optical response of the NINJA FROST detector described in Section~\ref{sec:simulation}, together with the relevant surrounding materials in the detector hall.

For the evaluation of the muon position reconstruction, events in which the muon was detected by BabyMIND were selected.
In addition, the muon crossing position in the NINJA FROST detector was required to be at least $10~\mathrm{mm}$ away from the detector edges to exclude regions with degraded position resolution.
When multiple position candidates were reconstructed, the candidate closest to the true muon position was taken as the reconstructed muon position.
Therefore, the resolution evaluated in this study represents the position resolution assuming that the muon candidate is correctly selected and does not include ambiguities in candidate selection.

\begin{figure}[htbp]
  \centering
  \includegraphics[width=1.0\linewidth]{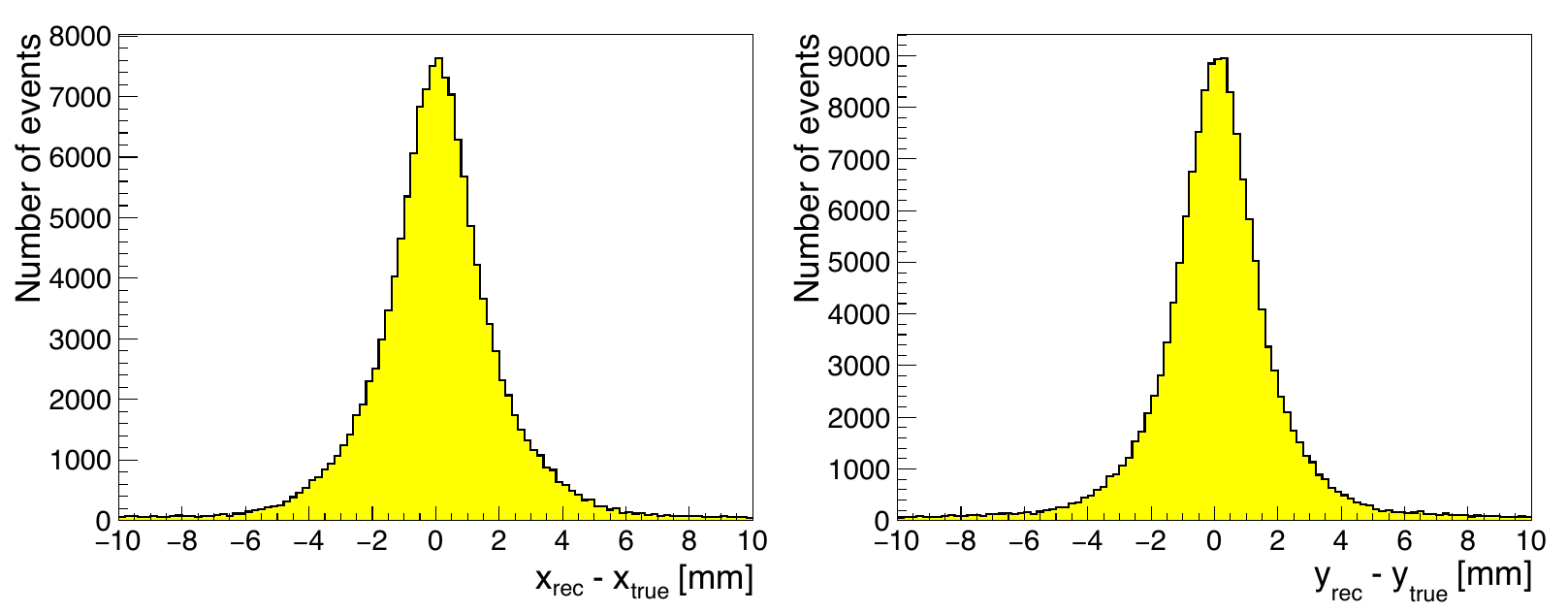}
  \caption{Distributions of $x_{\mathrm{rec}}-x_{\mathrm{true}}$ (left) and $y_{\mathrm{rec}}-y_{\mathrm{true}}$ (right) for selected muons in the neutrino-interaction MC sample.}
  \label{fig:muon_residuals}
\end{figure}

\begin{figure}[htbp]
  \centering
  \includegraphics[width=1.0\linewidth]{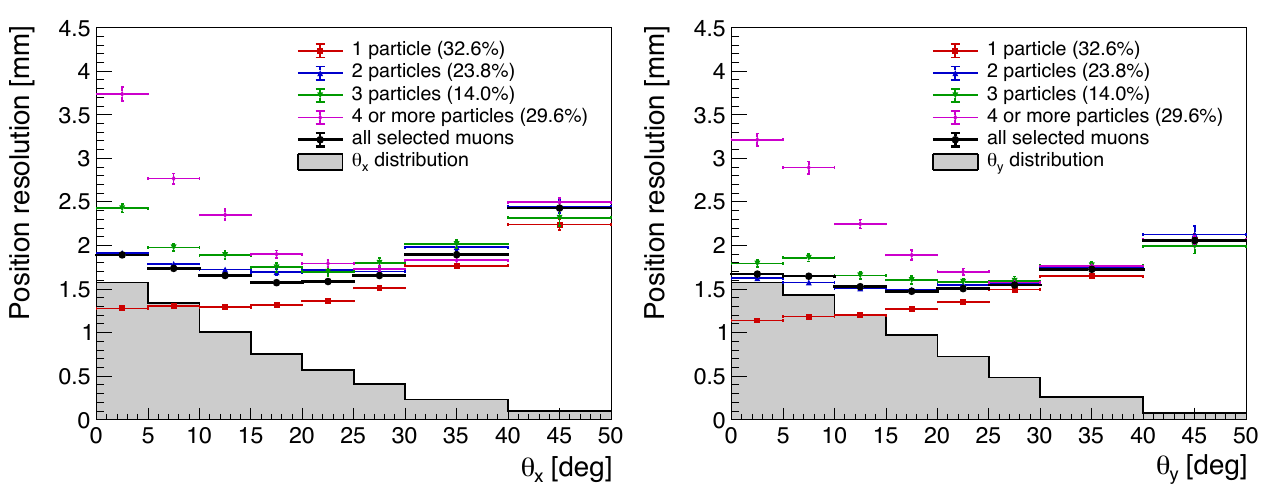}
  \caption{Position resolution in the $x$ (left) and $y$ (right) directions as a function of $\theta_x$ and $\theta_y$, respectively, for all selected muons and for different truth-level particle multiplicities. The percentages in parentheses indicate the fraction of selected events in each particle-multiplicity category. The shaded histograms show the corresponding muon angular distributions.}
  \label{fig:muon_resolution_angle}
\end{figure}

Figure~\ref{fig:muon_residuals} shows the residual distributions,
$x_{\mathrm{rec}}-x_{\mathrm{true}}$ and
$y_{\mathrm{rec}}-y_{\mathrm{true}}$,
for the selected muons.
Because the residual distributions are not necessarily Gaussian, they were characterized by their median $m$ and the half-width $\sigma_{68}$ of the central 68\% interval.
The position resolution was defined as
\begin{equation}
    \sigma_{\mathrm{pos}}
    =
    \sqrt{m^2+\sigma_{68}^2}.
    \label{eq:position_resolution_simulation}
\end{equation}
This definition accounts for both the spread of the residual distribution and any bias of its center from zero.
The resulting position resolutions were $1.77~\mathrm{mm}$ in the $x$ direction and $1.61~\mathrm{mm}$ in the $y$ direction, both well below the required position resolution of $5.6~\mathrm{mm}$.
The slightly better position resolution in the $y$ direction is attributed to the shorter WLS fibers used for the $y$ readout, which reduce light attenuation and result in a larger collected light yield.
These values are also much smaller than the $10~\mathrm{mm}/\sqrt{12}=2.89~\mathrm{mm}$ expected for a purely segmented detector with a $10~\mathrm{mm}$ segment width and a uniform distribution of particle crossing positions within each segment.

Figure~\ref{fig:muon_resolution_angle} shows the position resolution in the $x$ and $y$ directions as a function of the projected muon angles $\theta_x$ and $\theta_y$, respectively, for all selected muons and separately for truth-level particle multiplicities of one, two, three, and four or more.
The angles $\theta_x$ and $\theta_y$ were calculated from the true muon track directions in the $x$--$z$ and $y$--$z$ planes, respectively, taking the $z$ axis as the reference ($0^\circ$) direction.

For single-particle events, the position resolution degraded with increasing projected angle.
This degradation is caused by inclined muons traversing a wider range of positions in the scintillator and by the angle dependence of the weighted center of light yield relative to the normal-incidence mapping function.
For events with two or more particles, the position resolution was worse at small angles, reached its best values at approximately $20^\circ$--$30^\circ$, and degraded again at larger angles.
The poorer position resolution at small angles is caused by other particles tending to cross the detector near the muon crossing position, whereas the degradation at large angles arises from the same effects as in single-particle events.
For all particle multiplicities, the position resolution remained below the required $5.6~\mathrm{mm}$ over the entire angular range considered.

\section{Operation and data acquisition stability}
\label{sec:operation}

\subsection{Operating conditions}

The NINJA FROST detector was operated throughout the neutrino-beam exposure during the third NINJA physics run from November 2025 to March 2026.
During the exposure, the detector was installed in the NINJA detector rack, which was enclosed with thermal insulation and cooled.
The temperature around the detector was maintained at approximately $10^\circ\mathrm{C}$ to maintain stable nuclear-emulsion sensitivity and SiPM response.

\subsection{Periodic SiPM gain calibration}

Because only a small fraction of channels contained particle-induced signals in each beam-triggered event, the remaining channels primarily recorded SiPM dark noise.
These dark-noise data were used to determine the pedestal and single-photoelectron gain for each channel, as described in Section~\ref{sec:daq}.
The calibration was performed at intervals of 10,000 beam-triggered events, corresponding to approximately $3.6~\mathrm{h}$ of operation.
The temperature variation within each calibration interval was less than $0.3^\circ\mathrm{C}$.

\subsection{Beam exposure and data acquisition stability}

Data were collected with the J-PARC neutrino beam in both the $\nu_\mu$-focused and $\bar{\nu}_\mu$-focused modes~\cite{JNUBEAM}.
The $\nu_\mu$-focused data were collected from November 29 to December 22, 2025, from January 17 to January 28, 2026, and from March 11 to March 25, 2026, while the $\bar{\nu}_\mu$-focused data were collected from January 28 to March 11, 2026.
The delivered POT (protons on target) and the POT corresponding to the data recorded by the detector are summarized in Table~\ref{tab:operation_pot}.
The data acquisition efficiency was defined as the ratio of the recorded POT to the delivered POT.
In total, the detector recorded data corresponding to $1.0217\times10^{21}$ POT out of $1.0259\times10^{21}$ POT delivered during the beam exposure, corresponding to a data acquisition efficiency of $99.59\%$.

\begin{table}[htbp]
  \centering
  \caption{Summary of the beam exposure and data acquisition.}
  \label{tab:operation_pot}
  \begin{tabular}{cccc}
    \hline\hline
    Beam mode & Delivered POT ($10^{20}$) & Recorded POT ($10^{20}$) & Efficiency \\
    \hline
    $\nu_\mu$-focused   & 4.741  & 4.714  & 99.43\% \\
    $\bar{\nu}_\mu$-focused   & 5.518  & 5.503  & 99.73\% \\
    Total & 10.259 & 10.217 & 99.59\% \\
    \hline\hline
  \end{tabular}
\end{table}

\begin{figure}[htbp]
  \centering
  \includegraphics[width=1.0\linewidth]{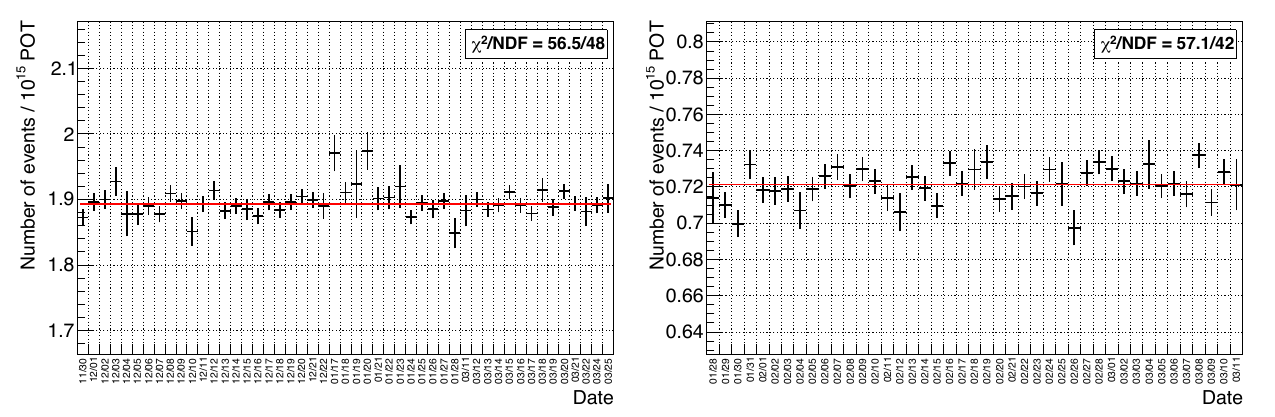}
  \caption{Daily event rate per recorded POT during the $\nu_\mu$-focused beam mode (left) and $\bar{\nu}_\mu$-focused beam mode (right).
  The red lines indicate fits assuming a constant event rate.}
  \label{fig:operation_event_rate}
\end{figure}

The stability of the detector operation was monitored using the number of selected events per recorded POT for each day.
Events were selected by requiring the maximum light yield to exceed $20~\mathrm{p.e.}$ in both readout directions within a beam bunch.
Most of the selected events were ``sand muons,'' produced by neutrino interactions in upstream material such as the surrounding sand and rock.
The number of such events is therefore expected to be proportional to the POT, provided that event pileup is negligible.
Figure~\ref{fig:operation_event_rate} shows the event rate during the $\nu_\mu$-focused and $\bar{\nu}_\mu$-focused beam modes.
The error bars represent statistical uncertainties only.
Fits assuming a constant event rate gave $\chi^2/\mathrm{NDF}=56.5/48$ for the $\nu_\mu$-focused mode and $57.1/42$ for the $\bar{\nu}_\mu$-focused mode.
The event rate remained stable throughout the beam exposure in both beam modes.

\section{Performance evaluation using beam data}
\label{sec:performance}

The performance of the NINJA FROST detector was evaluated in terms of the muon detection efficiency and the position resolution using sand muons recorded during the beam exposure.
A corresponding sand-muon MC sample was used to interpret the detection-efficiency measurement and to evaluate the position resolution through data--MC comparison.
The sand-muon MC sample was generated using the simulation framework described in Section~\ref{sec:reconstruction_simulation}, with neutrino interactions in the surrounding materials simulated with NEUT and the resulting particles propagated through the detector geometry with Geant4.

\subsection{Muon detection efficiency}
\label{sec:hit_efficiency}

The muon detection efficiency was evaluated for selected muon tracks expected to traverse the NINJA FROST detector.
The efficiency was defined as the fraction of these tracks for which a hit was observed in the detector.
A hit was considered to be observed when the maximum light yield was at least $10~\mathrm{p.e.}$ in both the $x$ and $y$ readout directions.
To select muons expected to pass through the detector, sand-muon events with a reconstructed BabyMIND track penetrating the Proton Module were selected.
As shown in Fig.~\ref{fig:ninja_wagasci_setup}, the Proton Module is located upstream of the detector, while BabyMIND is located downstream.
To ensure that the selected muons passed through the active area of the detector, the extrapolated BabyMIND track position at the detector plane was required to be at least $300~\mathrm{mm}$ from the detector edges in the $x$ direction and $200~\mathrm{mm}$ in the $y$ direction.
The smaller margin in the $y$ direction reflects the better BabyMIND track-extrapolation precision in that direction due to its better angular resolution.

\begin{figure}[htbp]
  \centering
  \includegraphics[width=0.6\linewidth]{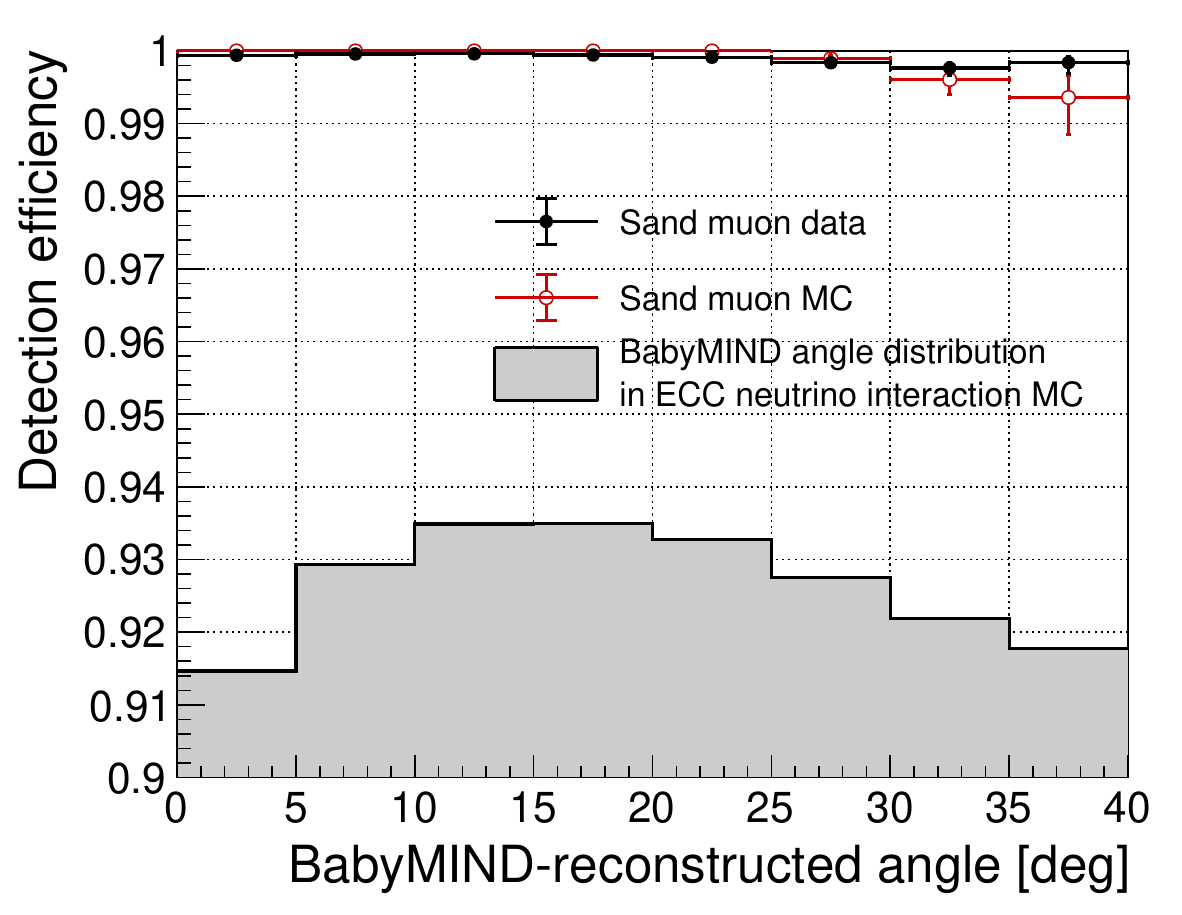}
  \caption{FROST muon detection efficiency as a function of the track angle reconstructed in BabyMIND for sand-muon data and MC. The shaded histogram shows the corresponding BabyMIND-reconstructed angle distribution of muons from simulated neutrino interactions in the ECCs.}
  \label{fig:hit_efficiency}
\end{figure}

Figure~\ref{fig:hit_efficiency} shows the detection efficiency as a function of the track angle reconstructed in BabyMIND for sand-muon data and MC.
The total detection efficiencies were $99.946\pm0.006\%$ in data and $99.989\pm0.004\%$ in MC, where the quoted uncertainties are statistical.
A small decrease in the detection efficiency was observed at large reconstructed angles.
This decrease is caused by the limited accuracy of the BabyMIND track extrapolation, which becomes worse at large angles.
To confirm this interpretation, the events contributing to the inefficiency in MC were examined using the true muon trajectories.
All such events were found to correspond to muons that did not actually cross the detector, although their extrapolated BabyMIND track positions satisfied the selection criterion.
No inefficiency was observed in the MC sample for muons that actually crossed the detector.

\subsection{Upper limit on the position resolution using a single-muon-dominated sample}
\label{sec:position_resolution_data}

The position resolution of the NINJA FROST detector was evaluated using a sand-muon sample, in which almost all selected events contained a single muon.
The position resolution for multi-particle events was not evaluated using beam data because a sufficiently large data sample was not available.

\subsubsection{Reference-track reconstruction and event selection}

A reference muon track was reconstructed using hits in the Proton Module (PM) and Downstream WAGASCI (DWG), without using information from the NINJA FROST detector.
As shown in Fig.~\ref{fig:ninja_wagasci_setup}, the PM is located upstream of the detector, while DWG is located immediately downstream.
A straight line was fitted to the hits in the PM and DWG and extrapolated to the detector plane.
The extrapolated positions in the $x$ and $y$ directions are defined as $x_{\mathrm{ext}}$ and $y_{\mathrm{ext}}$, respectively.
The corresponding track angles measured from the $z$ axis in the $x$--$z$ and $y$--$z$ planes are denoted by $\theta_x$ and $\theta_y$, respectively.

Sand-muon events with a single reconstructed muon track in BabyMIND per spill were selected.
The reference track was required to contain at least 5 DWG hits and 10 PM hits in each of the $x$ and $y$ projections to ensure sufficient accuracy of the track fit.
The reconstructed position in the detector was also required to be at least $10~\mathrm{mm}$ from the detector edges in both directions to exclude regions with degraded position resolution, as in the simulation study described in Section~\ref{sec:reconstruction_simulation}.

The position resolution of the extrapolated reference position was approximately $4~\mathrm{mm}$, which was larger than the position resolution expected from MC for the NINJA FROST detector.
Therefore, the widths of the $x_{\mathrm{rec}}-x_{\mathrm{ext}}$ and $y_{\mathrm{rec}}-y_{\mathrm{ext}}$ distributions could not be directly interpreted as the detector position resolution.
Instead, the residual distributions in data and MC were compared to estimate an upper limit on the position resolution.
The evaluation was performed separately for $0^\circ\leq\theta<10^\circ$ and $10^\circ\leq\theta<20^\circ$ in each direction.
A bin width of $10^\circ$ was chosen to account for the angular dependence while maintaining sufficient precision of the PM--DWG reference-track extrapolation.
The upper angular limit of $20^\circ$ was determined by the geometrical acceptance for sand muons traversing the PM, the NINJA FROST detector, and DWG.

\subsubsection{Data--MC comparison and reference-track validation}

\begin{figure}[htbp]
  \centering
  \includegraphics[width=1.0\linewidth]{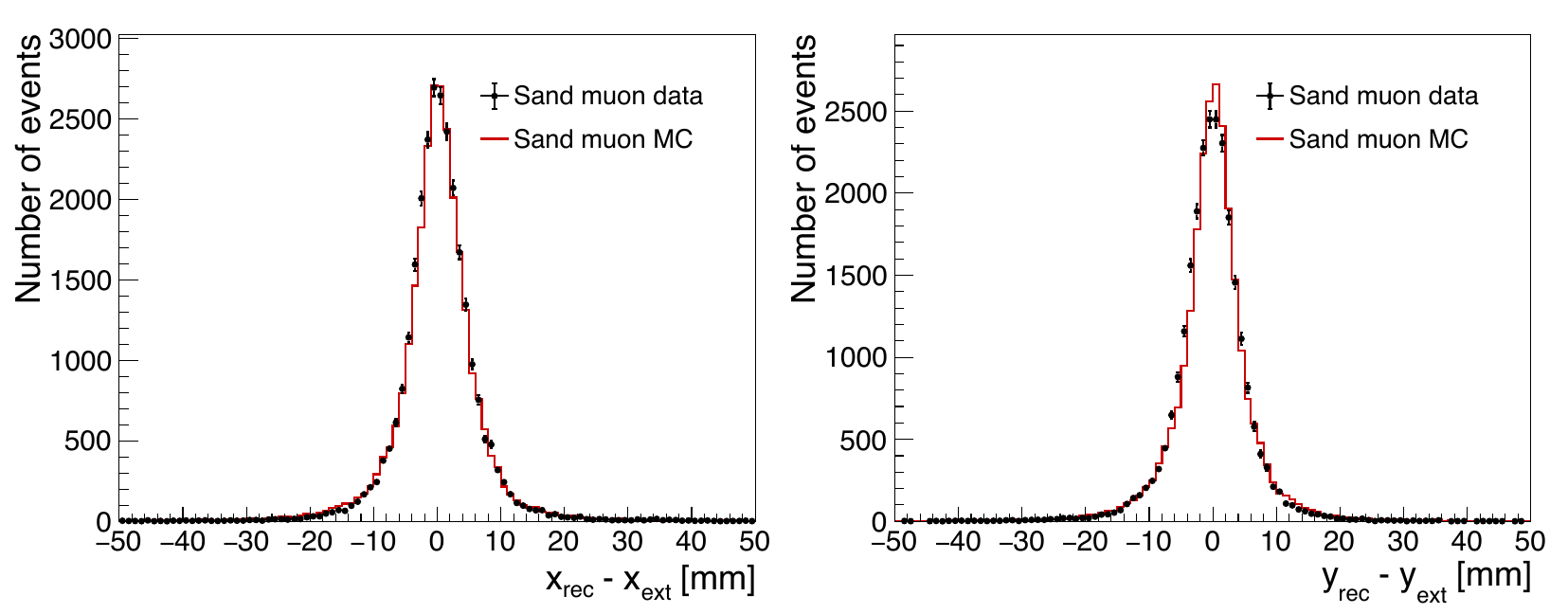}
  \caption{Distributions of $x_{\mathrm{rec}}-x_{\mathrm{ext}}$ for $0^\circ\leq\theta_x<10^\circ$ (left) and $y_{\mathrm{rec}}-y_{\mathrm{ext}}$ for $0^\circ\leq\theta_y<10^\circ$ (right) in sand-muon data and MC.}
  \label{fig:position_resolution_residuals}
\end{figure}

Figure~\ref{fig:position_resolution_residuals} shows the residual distributions in data and MC for
$0^\circ\leq\theta_x<10^\circ$ in the $x$ direction and
$0^\circ\leq\theta_y<10^\circ$ in the $y$ direction.
The widths of the distributions were characterized by $\sigma_{68}$, defined as the half-width of the central 68\% interval.
In the following, $\sigma_{68}(r)$ denotes the $\sigma_{68}$ value for the distribution of residual $r$.
The data--MC comparison was performed for both angular ranges in each direction.
For each angular range, the absolute relative difference in the $x$ direction was calculated as
\begin{equation}
    R
    =
    \left|
    \frac{
        \sigma_{68}^{\mathrm{data}}(x_{\mathrm{rec}}-x_{\mathrm{ext}})
        -
        \sigma_{68}^{\mathrm{MC}}(x_{\mathrm{rec}}-x_{\mathrm{ext}})
    }{
        \sigma_{68}^{\mathrm{MC}}(x_{\mathrm{rec}}-x_{\mathrm{ext}})
    }
    \right|.
    \label{eq:data_mc_difference}
\end{equation}
The relative difference in the $y$ direction was calculated in the same manner.
The largest value among the four comparisons was $4.4\%$, which was taken as $R_{\max}$.

The residual width contains contributions from both the reconstructed position in the detector and the extrapolated reference position, and the data--MC difference can therefore arise from either contribution.
To independently check the data--MC agreement relevant to the reference-track extrapolation, the PM hits and DWG hits were fitted separately.
The positions obtained from the PM-only and DWG-only fits were independently extrapolated to the NINJA FROST detector plane, and the width of the difference between these two extrapolated positions was compared between data and MC.
The relative differences in the $\sigma_{68}$ values of these position-difference distributions ranged from $-1.4\%$ to $+6.4\%$ among the two projections and two angular ranges, with no systematic tendency for the widths in data to be smaller than those in MC.
When the two angular ranges were combined, the relative differences were $+2.4\%$ and $-0.8\%$ in the $x$ and $y$ directions, respectively.
Thus, the comparison showed no indication that the reference-track resolution was substantially better in data than in MC.

\subsubsection{Estimation of the position-resolution upper limit}

In light of this comparison, an upper-limit estimate of the FROST position resolution was derived under the assumption that the reference-track contribution to the residual width was not smaller in data than in MC.

The difference between the squared residual widths in data and MC was defined as
\begin{equation}
  \Delta\sigma_{68}^{2}
  (x_{\mathrm{rec}}-x_{\mathrm{ext}})
  \equiv
  \left\{
      \sigma_{68}^{\mathrm{data}}
      (x_{\mathrm{rec}}-x_{\mathrm{ext}})
  \right\}^{2}
  -
  \left\{
      \sigma_{68}^{\mathrm{MC}}
      (x_{\mathrm{rec}}-x_{\mathrm{ext}})
  \right\}^{2}.
  \label{eq:residual_width_difference}
\end{equation}
Since the absolute relative difference in $\sigma_{68}$ was at most $R_{\max}$, the magnitude of the difference in the squared residual width is bounded by
\begin{equation}
  \left|
      \Delta\sigma_{68}^{2}
      (x_{\mathrm{rec}}-x_{\mathrm{ext}})
  \right|
  \leq
  \left\{(1+R_{\max})^{2}-1\right\}
  \left\{
      \sigma_{68}^{\mathrm{MC}}
      (x_{\mathrm{rec}}-x_{\mathrm{ext}})
  \right\}^{2}.
  \label{eq:additional_residual_width}
\end{equation}

For the upper-limit estimate, the positive bound in Eq.~\eqref{eq:additional_residual_width} was used as a possible increase in the squared residual width, which was attributed entirely to FROST.
The corresponding upper-limit $\sigma_{\mathrm{pos,UL}}^{x,\mathrm{data}}$ was therefore defined by
\begin{equation}
  \left(\sigma_{\mathrm{pos,UL}}^{x,\mathrm{data}}\right)^2
  -
  \left(\sigma_{\mathrm{pos}}^{x,\mathrm{MC}}\right)^2
  =
  \left\{(1+R_{\max})^{2}-1\right\}
  \left\{
      \sigma_{68}^{\mathrm{MC}}
      (x_{\mathrm{rec}}-x_{\mathrm{ext}})
  \right\}^{2}.
  \label{eq:position_resolution_increase}
\end{equation}
Here, the detector position resolution in MC was evaluated using the definition in Eq.~\eqref{eq:position_resolution_simulation}, as
\begin{equation}
  \left(\sigma_{\mathrm{pos}}^{x,\mathrm{MC}}\right)^2
  =
  \left\{
      m^{\mathrm{MC}}
      (x_{\mathrm{rec}}-x_{\mathrm{true}})
  \right\}^{2}
  +
  \left\{
      \sigma_{68}^{\mathrm{MC}}
      (x_{\mathrm{rec}}-x_{\mathrm{true}})
  \right\}^{2},
  \label{eq:mc_position_resolution_x}
\end{equation}
where $m^{\mathrm{MC}}(x_{\mathrm{rec}}-x_{\mathrm{true}})$ is the median of the residual distribution in MC.
The upper limit in the $y$ direction was evaluated in the same manner.

The results are summarized in Table~\ref{tab:position_resolution_upper_limit}.
All upper limits were below the required position resolution of $5.6~\mathrm{mm}$ (see Section~\ref{sec:frost_requirements}) and the $10~\mathrm{mm}/\sqrt{12}=2.89~\mathrm{mm}$ expected for a purely segmented detector with a $10~\mathrm{mm}$ segment width.
They were also smaller than the position resolution of the previous scintillator tracker, which varied from approximately $2.4$ to $3.8~\mathrm{mm}$ depending on the track angle for angles below $20^\circ$~\cite{E71aTracker}.

\begin{table}[htbp]
  \centering
  \caption{Position-resolution results for sand muons. The $\sigma_{68}$ values are obtained from the $x_{\mathrm{rec}}-x_{\mathrm{ext}}$ or $y_{\mathrm{rec}}-y_{\mathrm{ext}}$ residual distributions.}
  \label{tab:position_resolution_upper_limit}
  \begin{tabular}{cccccc}
    \hline\hline
    Direction & Angle
    & $\sigma_{68}^{\mathrm{data}}$~[mm]
    & $\sigma_{68}^{\mathrm{MC}}$~[mm]
    & $\sigma_{\mathrm{pos}}^{\mathrm{MC}}$~[mm]
    & $\sigma_{\mathrm{pos,UL}}^{\mathrm{data}}$~[mm] \\
    \hline
    $x$ & $0^\circ$--$10^\circ$  & 4.54 & 4.70 & 1.43 & 2.01 \\
    $x$ & $10^\circ$--$20^\circ$ & 5.36 & 5.61 & 1.43 & 2.20 \\
    $y$ & $0^\circ$--$10^\circ$  & 4.40 & 4.39 & 1.32 & 1.87 \\
    $y$ & $10^\circ$--$20^\circ$ & 4.93 & 5.07 & 1.35 & 2.03 \\
    \hline\hline
  \end{tabular}
\end{table}

\section{Conclusion}
\label{sec:conclusion}

In this paper, we presented the construction, position reconstruction, operation, and performance of a meter-scale FROST detector used in the NINJA experiment.
The NINJA FROST detector consists of a $1.32~\mathrm{m}\times1.40~\mathrm{m}$ plastic scintillator plate assembled from optically bonded scatterer-embedded plates, with WLS fibers arranged in two orthogonal directions at a pitch of $10~\mathrm{mm}$ and 272 readout channels in total.
The collected light yield tends to be larger in channels whose fibers are closer to the particle crossing position, and the embedded scatterers further strengthen this tendency by localizing the scintillation light around the crossing position.
The crossing position can therefore be reconstructed from the light-yield distribution across the readout channels with a position resolution well below the readout pitch.

A position reconstruction method capable of handling both single- and multi-particle events was developed.
In the neutrino-interaction MC, the muon position resolutions for the selected sample were $1.77~\mathrm{mm}$ in the $x$ direction and $1.61~\mathrm{mm}$ in the $y$ direction assuming that the muon candidate is correctly selected, both well below the $5.6~\mathrm{mm}$ position resolution required for the scintillator tracker in the NINJA experiment.

The detector was operated throughout the neutrino-beam exposure and recorded data corresponding to $1.0217\times10^{21}$ POT with a data acquisition efficiency of $99.59\%$.
The event rate per recorded POT remained stable throughout the exposure.

The detector performance was evaluated using sand-muon data recorded during the beam exposure, which were dominated by single-muon events.
The measured detection efficiency for selected muons well within the active area was $99.946\pm0.006\%$.
For track angles below $20^\circ$, the upper limits on the position resolution were at most $2.20~\mathrm{mm}$ in the $x$ direction and $2.03~\mathrm{mm}$ in the $y$ direction.
All upper limits were below the required $5.6~\mathrm{mm}$ and the $2.89~\mathrm{mm}$ expected for a purely segmented detector with a $10~\mathrm{mm}$ segment width.
These results demonstrate that the FROST concept can be scaled to a meter-scale active area while maintaining high detection efficiency and a position resolution well below the readout pitch under actual experimental conditions.

\section*{CRediT authorship contribution statement}

\textbf{N. Otani}: Conceptualization, Methodology, Software, Validation, Formal analysis, Investigation, Writing -- original draft, Writing -- review \& editing, Visualization, Funding acquisition.
\textbf{S. Han}: Software, Investigation, Writing -- review \& editing.
\textbf{E. Abad D\'{i}az}: Investigation.
\textbf{S. Aoki}: Writing -- review \& editing.
\textbf{D. Bar\v{c}ot}: Investigation.
\textbf{C. Bronner}: Investigation, Writing -- review \& editing.
\textbf{T. Fukuda}: Investigation.
\textbf{M. Ghosh}: Investigation.
\textbf{L. Hali\'{c}}: Investigation.
\textbf{K. Hasegawa}: Investigation.
\textbf{T. Hayakawa}: Investigation.
\textbf{Y. Hayasaka}: Investigation.
\textbf{K. Hayashi}: Investigation.
\textbf{D. Hirata}: Software, Formal analysis, Investigation, Resources.
\textbf{K. Indo}: Investigation.
\textbf{M. Jakkapu}: Investigation.
\textbf{A. Kasumi}: Investigation.
\textbf{T. Katori}: Investigation, Writing -- review \& editing.
\textbf{T. Kawahara}: Investigation.
\textbf{M. Kawaue}: Investigation, Writing -- review \& editing.
\textbf{T. Kikawa}: Conceptualization, Investigation, Writing -- review \& editing, Supervision, Project administration, Funding acquisition.
\textbf{B. Kli\v{c}ek}: Investigation.
\textbf{H. Kobayashi}: Investigation.
\textbf{B. Kova\v{c}}: Investigation.
\textbf{J. McKean}: Investigation, Writing -- review \& editing.
\textbf{T. Matsuo}: Investigation.
\textbf{S. Mikado}: Investigation.
\textbf{A. Minamino}: Investigation.
\textbf{Y. Morimoto}: Investigation.
\textbf{Y. Nakamura}: Investigation.
\textbf{T. Nakaya}: Conceptualization, Investigation, Writing -- review \& editing, Supervision, Project administration, Funding acquisition.
\textbf{H. Oaira}: Investigation.
\textbf{N. Onda}: Investigation, Writing -- review \& editing.
\textbf{H. Oshima}: Investigation.
\textbf{O. Sato}: Investigation.

\section*{Declaration of generative AI and AI-assisted technologies in the manuscript preparation process}

During the preparation of this work, the authors used ChatGPT in order to improve the readability and language of the manuscript. After using this tool, the authors reviewed and edited the content as needed and take full responsibility for the content of the published article.

\section*{Acknowledgments}

We appreciate the support of the T2K Collaboration in conducting the NINJA experiment and thank the J-PARC staff for the stable operation of the accelerator.
We thank the T2K neutrino beam group for delivering a stable neutrino beam and for their support with the neutrino beam simulation.
We acknowledge the WAGASCI/BabyMIND group for the stable operation of WAGASCI and BabyMIND and for making the experimental data and software available.
We gratefully acknowledge the SPADI Alliance Working Group 1 RAYRAW Task Force for providing us with the opportunity to use the RAYRAW-v1 readout board during its development phase.
We would also like to express our special thanks to Ryotaro Honda and Rintaro Kurata for their extensive technical support and valuable advice on the use of the RAYRAW-v1 readout board throughout this work.
We are grateful to Kuraray Co., Ltd. for fabricating the custom scatterer-embedded plastic scintillator used in the NINJA FROST detector.

This work was supported by JSPS KAKENHI Grant Numbers JP23H05434, JP24K00644, JP24K23938, and JP25KJ1470, and was carried out by the joint research program of the Institute of Materials and Systems for Sustainability (IMaSS), Nagoya University.
It was also supported by the Croatian Science Foundation under the project numbers HRZZ-DOK-NPOO-2023-10-1262 and HRZZ-DOK-2025-02-2372; Swiss National Science Foundation and Croatian Science Foundation via grant MAPS IZ11Z0\_230193; Ministry of Science, Education and Youth of Republic of Croatia via grant No. PK.1.1.10.0002 and via financial support for participating in the international NINJA project.

\end{document}